\documentstyle [12pt]{article}

\newcommand{\beq}{\begin{equation}}
\newcommand{\eeq}{\end{equation}}

\def\a{\alpha}
\def\b{\beta}

\def\d{\delta}
\def\e{\epsilon}

\def\g{\gamma}
\def\h{\eta}

\def\l{\lambda}
\def\m{\mu}

\def\o{\omega}

\def\t{\theta}
\def\r{\rho}
\def\s{\sigma}

\def\G{\Gamma}

\def\L{\Lambda}
\def\O{\Omega}

\def\S{\Sigma}

\def\bfh{\mbox{\boldmath $\h$}}

\def\bfo{\mbox{\boldmath $\o$}}

\def\bft{\mbox{\boldmath $\t$}}

\def\bfpsi{\mbox{\boldmath $\psi$}}

\def\bo{{\raise.15ex\hbox{\large$\Box$}}}
\def\pa{\partial}

\def\TH{{\raise.2ex\hbox{$\displaystyle \bigodot$}\mskip-4.7mu \llap H \;}}
\def\face{{\raise.2ex\hbox{$\displaystyle \bigodot$}\mskip-2.2mu \llap {$\ddot
        \smile$}}}

\def\slash#1{\rlap/#1}

\def\Hat#1{\rlap{\kern.10em$\widehat{\phantom G}$}#1}
\def\HAt#1{\rlap{\kern.05em$\widehat{\phantom G}$}#1}

\def\cap#1{\rlap{\kern.1em$\widehat{\phantom{G\vrule height.8em}}$}#1{}}
\def\Cap#1{\rlap{\kern.05em$\widehat{\phantom{G\vrule height.8em}}$}#1{}}

\def\leftrightarrowfill{$\mathsurround=0pt \mathord\leftarrow \mkern-6mu
        \cleaders\hbox{$\mkern-2mu \mathord- \mkern-2mu$}\hfill
        \mkern-6mu \mathord\rightarrow$}
\def\overleftrightarrow#1{\vbox{\ialign{##\crcr
        \leftrightarrowfill\crcr\noalign{\kern-1pt\nointerlineskip}
        $\hfil\displaystyle{#1}\hfil$\crcr}}}

\def\sfrac#1#2{{\vphantom1\smash{\lower.5ex\hbox{\small$#1$}}\over
        \vphantom1\smash{\raise.4ex\hbox{\small$#2$}}}}
\def\bfrac#1#2{{\vphantom1\smash{\lower.5ex\hbox{$#1$}}\over
        \vphantom1\smash{\raise.3ex\hbox{$#2$}}}}
\def\afrac#1#2{{\vphantom1\smash{\lower.5ex\hbox{$#1$}}\over#2}}

\catcode`@=11
\def\underline#1{\relax\ifmmode\@@underline#1\else
        $\@@underline{\hbox{#1}}$\relax\fi}
\catcode`@=12

\def\nis{\nointerlineskip}
\def\Abar{\vbox{\nis\moveright.33em\vbox{
        \hrule width.35em height.04em}\nis\kern.05em\hbox{$A$}}{}}
\def\Dbar{\vbox{\nis\moveright.20em\vbox{
        \hrule width.50em height.04em}\nis\kern.05em\hbox{$D$}}{}}
\def\Gbar{\vbox{\nis\moveright.20em\vbox{
        \hrule width.50em height.04em}\nis\kern.05em\hbox{$G$}}{}}
\def\mbar{\vbox{\nis\moveright.15em\vbox{
        \hrule width.60em height.04em}\nis\kern.05em\hbox{$m$}}{}}
\def\Rbar{\vbox{\nis\moveright.20em\vbox{
        \hrule width.50em height.04em}\nis\kern.05em\hbox{$R$}}{}}
\def\Vbar{\vbox{\nis\moveright.05em\vbox{
        \hrule width.60em height.04em}\nis\kern.05em\hbox{$V$}}{}}
\def\Xbar{\vbox{\nis\moveright.20em\vbox{
        \hrule width.60em height.04em}\nis\kern.05em\hbox{$X$}}{}}
\def\thetabar{\vbox{\nis\moveright.15em\vbox{
        \hrule width.30em height.04em}\nis\kern.05em\hbox{$\theta$}}{}}
\def\Lambdabar{\vbox{\nis\moveright.25em\vbox{
        \hrule width.35em height.04em}\nis\kern.05em\hbox{${\mit\Lambda}$}}{}}
\def\Sigmabar{\vbox{\nis\moveright.25em\vbox{
        \hrule width.50em height.04em}\nis\kern.05em\hbox{${\mit\Sigma}$}}{}}
\def\phibar{\vbox{\nis\moveright.18em\vbox{
        \hrule width.40em height.04em}\nis\kern.05em\hbox{$\phi$}}{}}
\def\chibar{\vbox{\nis\moveright.12em\vbox{
        \hrule width.40em height.04em}\nis\kern.05em\hbox{$\chi$}}{}}
\def\psibar{\vbox{\nis\moveright.23em\vbox{
        \hrule width.40em height.04em}\nis\kern.05em\hbox{$\psi$}}{}}
\def\debar{\vbox{\nis\moveright.18em\vbox{
        \hrule width.35em height.04em}\nis\kern.05em\hbox{$\partial$}}{}}
\def\delbar{\vbox{\nis\moveright.10em\vbox{
        \hrule width.63em height.04em}\nis\kern.05em\hbox{$\nabla$}}{}}

\newskip\humongous \humongous=0pt plus 1000pt minus 1000pt

\newif\ifdtup

\begin{document}
\title{\bf On the eigenfunctions of the Dirac operator on spheres and
real hyperbolic spaces}
\author{Roberto Camporesi\\
{\it Dipartimento di Matematica, Politecnico di Torino}\\
{\it Corso Duca degli Abruzzi 24, 10129 Torino Italy}
\\and\\
Atsushi Higuchi\\
{\it Institut f\"ur theoretische Physik, Universit\"at Bern}\\
{\it Sidlerstrasse 5, CH-3012 Bern, Switzerland}}
\maketitle
\begin{abstract}
The eigenfunctions
of the Dirac operator on spheres and real hyperbolic spaces of
arbitrary dimension are computed by separating variables in geodesic polar
coordinates. These eigenfunctions are used to derive the heat kernel of the
iterated Dirac operator on these spaces.
They are then studied as cross sections of homogeneous vector bundles,
and a group-theoretic derivation of the spinor spherical functions and
heat kernel is given based on
Harish-Chandra's formula for the radial part of the Casimir
operator.
\end{abstract}

\newpage

\section{Introduction}

The $N$-dimensional sphere ($S^N$) and the real hyperbolic space ($H^N$),
which are ``dual" to each other as symmetric spaces \cite{HELGA2}, are
maximally symmetric.  This high degree of symmetry
allows one to compute explicitly the eigenfunctions of the
Laplacian
for various
fields on these spaces.
These eigenfunctions can be used in studying field theory
in de~Sitter and anti-de~Sitter spacetimes since $S^4$ and $H^4$ are Euclidean
sections of these spacetimes.
Also $S^3$ and $H^3$ appear
as the spatial sections of cosmological models, and various
field equations and their solutions
on these spaces have physical applications in this context.
In addition to these applications, fields on $S^N$ and $H^N$ provide
concrete examples for the theory of homogeneous vector bundles, and
consequences of various theorems can explicitly be worked out.

Recently the authors presented the eigenfunctions of the Laplacian for the
transverse-traceless totally-symmetric tensor
fields \cite{CAMPOHIGU} and for the
totally-antisymmetric tensor fields ($p$-forms) \cite{CAMPOHIGU2}.  These
fields were also analyzed
in the light of group theory using the fact that they
are cross sections of homogeneous vector bundles. As a continuation of these
works we study in this paper
the spinor fields satisfying the Dirac equation
$\slash{\! \nabla}\psi = i\l \psi$ and the heat kernel for
$\slash{\! \nabla}^2$ on $S^N$ and $H^N$.
The paper is organized as follows.  We begin by setting up the gamma matrices
used in the paper in sect.\ 2.
Then in sect.\ 3 the appropriately
normalized eigenfunctions of the Dirac operator on $S^N$ with arbitrary $N$
are presented
using geodesic polar coordinates.
The solutions on $S^N$ are
expressed in terms of those on $S^{N-1}$. Then we derive the degeneracies of
the Dirac operator using the spinor eigenfunctions.
Next, spinor eigenfunctions on $H^N$ are obtained by
analytically continuing those on $S^N$. Then they are used to
derive the spinor spectral function (Plancherel measure) on $H^N$.
In sect.\ 4 the results of sect.\ 3 are used to
write down the heat kernel for the iterated
Dirac operator $\slash{\! \nabla}^2$ on these spaces.
Sect.\ 5 is devoted to a group-theoretic analysis of spinor fields on
$S^N$ and $H^N$.
We use the fact that spinor fields on these symmetric spaces are cross
sections of the homogeneous vector bundles associated to the fundamental spinor
representation(s) of $Spin(N)$.
We first review some relevant facts about harmonic analysis for
homogeneous vector bundles over compact symmetric spaces.
By applying these  to
$S^N = Spin(N+1)/Spin(N)$, in combination with the
formula for the radial part of the Casimir operator given by
Harish-Chandra,
we derive the spinor spherical functions and the heat kernel of the
iterated Dirac operator on $S^N$. Then we briefly review harmonic analysis for
homogeneous vector bundles over noncompact symmetric spaces, apply it to
$H^N = Spin(N,1)/Spin(N)$, and rederive some
results of sect.\ 4 for this space.

\newpage

\section{$\G$-matrices in $N$ dimensions}

A Clifford algebra in $N$ dimensions is generated by $N$ matrices $\G^a$
 satisfying the anticommutation relations
\beq
\label{ACR}
\G^a\G^b+\G^b\G^a=2\d^{ab}{\bf 1},\;\;\;\;\;\;\;a,b=1,\ldots ,N,
\eeq
where ${\bf 1}$ is the unit matrix and $\d^{ab}$ is the Kronecker symbol.
It is well known that (\ref{ACR}) can be
satisfied by matrices of dimension $2^{\left[\frac{N}{2}\right]}$, where
$\left[\frac{N}{2}\right]=N/2$ for $N$ even,
$\left[\frac{N}{2}\right]=(N-1)/2$ for $N$ odd. We give below an
inductive construction of $\G^a$ which relates easily to the spinor
representations of the orthogonal groups.

\

For $N=1$ we set $\G^1=1\in {\bf R}$. For $N=2$ put
\beq
\label{IND1}
\G^2=\left( \begin{array}{cc}0&1 \\  1&0
\end{array}\right),
\;\;\;\;\;\G^1=\left( \begin{array}{cc}0&i \\
-i&0
\end{array}\right),
\eeq
where $i=\sqrt{-1}$. For $N=3$ we add to $\G^1$ and $\G^2$
above the matrix
\beq
\label{IND2}
\G^3=(-i)\G^1\G^2=
\left( \begin{array}{cc}1&0 \\  0&-1
\end{array}\right).
\eeq

For $N=4$ let
\beq
\label{IND3}
\G^4=\left( \begin{array}{cc}0&{\bf 1} \\ {\bf 1}&0
\end{array}\right),
\;\;\;\;\;\G^j=\left( \begin{array}{cc}0&i\tilde{\G}^j \\
-i\tilde{\G}^j&0
\end{array}\right),\;\;\;\;\;j=1,2,3,
\eeq
where ${\bf 1}$ is the $2\times 2$ unit matrix, and the matrices
$\tilde{\G}^j$ ($j=1,2,3$)
satisfy the Clifford algebra for $N=3$ and are given by the right-hand
sides of (\ref{IND1})-(\ref{IND2}).
For $N=5$ we add to
$\G^j$ $(j=1,\ldots ,4)$ given in (\ref{IND3}) the matrix
\beq
\G^5=(-i)^2\G^1\G^2\G^3\G^4=
\left( \begin{array}{cc}{\bf 1}&0 \\  0&{\bf -1}
\end{array}\right).
\eeq

The general pattern is now clear.

\

Case 1. $N$  even. Let $\G^a=(\G^j,\G^N)$
denote the following set of matrices of dimension $2^{N/2}$:
\beq
\label{PRIMA}
\G^N=\left( \begin{array}{cc}0&{\bf 1} \\ {\bf 1}&0
\end{array}\right),
\;\;\;\;\;\G^j=\left( \begin{array}{cc}0&i\tilde{\G}^j \\
-i\tilde{\G}^j&0
\end{array}\right),\;\;\;\;\;\;j=1,\ldots ,N-1,
\eeq
where ${\bf 1}$ is the unit matrix of dimension $2^{N/2-1}$, and
the matrices $\tilde{\G}^j$ (also of dimension $2^{N/2-1}$)
satisfy the
Clifford algebra relations in $N-1$ dimensions,
\beq
\label{LD}
\tilde{\G}^j\tilde{\G}^k+\tilde{\G}^k\tilde{\G}^j=2\d^{jk},\;\;\;\;\;\;
j,k=1,\ldots ,N-1.
\eeq
Then
(\ref{ACR}) is immediately verified.
This representation generalizes to $N$ even the ordinary spinor
representation
of gamma matrices in four dimensions, where all $\g_a$'s are
``off-diagonal"
and $\g_5$ is diagonal.

The
$\frac{1}{2}N(N\!-\!1)$ matrices
\beq
\label{GENE}
\S^{ab}\equiv \frac{1}{4}[\G^a,\G^b],\;\;\;\;\;\;a,b=1,\ldots ,N
\eeq
satisfy the $SO(N)$ commutation rules
\beq
\label{SONCR}
[\S^{ab},\S^{cd}]=\d^{bc}\S^{ad}\!-\!\d^{ac}\S^{bd}\!-\!\d^{bd}\S^{ac
}\!+\!
\d^{ad}\S^{bc},
\eeq
and generate a $2^{N/2}$-dimensional representation of
$Spin(N)$, the double cover of $SO(N)$ (for $N>2$ $Spin(N)$ is also the
universal covering group).
The
commutator of
$\S^{ab}$ and $\G^c$ is then
\beq
\label{BELLO}
[\S^{ab},\G^c]=\d^{bc}\G^a-\d^{ac}\G^b.
\eeq

It is easily seen that the matrix
$\bfh =\G^1\G^2\cdots \G^N$ anticommutes with each of the $\G^a$'s, and
commutes with each of the $Spin(N)$ generators $\S^{ab}$. It can be shown
by induction that
\beq
\bfh=i^{N/2}
\left( \begin{array}{cc}{\bf 1}&0 \\  0&{\bf -1}
\end{array}\right).
\eeq
Since $\bfh$ is nontrivial it follows that the representation $\tau$
of $Spin(N)$
with generators $\S^{ab}$ must be reducible. From (\ref{PRIMA}) we find
\beq
\label{PRIMA2}
\S^{ab}=
\left( \begin{array}{cc}\S^{ab}_+&0 \\  0&\S^{ab}_-
\end{array}\right),
\eeq
where $\S^{ab}_+$ and $\S^{ab}_-$ both satisfy (\ref{SONCR}) and are given by
\begin{eqnarray}
\label{LG}
\S^{jk}_{\pm}&=&\tilde{\S}^{jk}\equiv \frac{1}{4}[\tilde{\G}^j,\tilde{\G}^k],
\\
\label{LG2}
\S^{jN}_{\pm}&=&-\S^{Nj}_{\pm}=\pm \frac{i}{2}\tilde{\G}^j,
\;\;\;\;\;\;\;\;j,k=1,
\ldots ,N\!-\!1,
\end{eqnarray}

It can be shown
that the matrices $\S^{ab}_+$ and $\S^{ab}_-$ generate the
irreducible representations of $Spin(N)$ with highest weights $\tau_+$
and
$\tau_-$ respectively, where
\beq
\label{FWSR1}
\tau_{\pm}=(\frac{1}{2},\frac{1}{2},\ldots ,\frac{1}{2},\pm\frac{1}{2})
\eeq
in the standard Cartan-Weyl labeling scheme.
Since these weights are fundamental (see \cite{BARUT} p.224), $\tau_+$
and
$\tau_-$ are called the two fundamental spinor representations of
$Spin(N)$ (of dimension $2^{N/2-1}$ each).
Thus $\tau=\tau_+\oplus \tau_-$ and a spinor in
$N$ dimensions ($N$ even) is reducible with respect to the orthogonal group.

In a similar way, the matrices $\tilde{\S}^{jk}$ given by (\ref{LG})
generate the unique
fundamental spinor representation $\s$ of $Spin(N\!-\!1)$
(see Case 2 below), also of dimension $2^{N/2-1}$, and one has
the branching rule
\beq
\tau_+|_{Spin(N\!-\!1)}=\s=\tau_-|_{Spin(N\!-\!1)}.
\eeq

\

Case 2. $N$ odd. In this case the dimension of the $\G$-matrices is
$2^{(N-1)/2}$, the same as in $N\!-\!1$ dimensions.
Let $\tilde{\G}^j$ ($j=1,\ldots ,N\!-\!1$) be a set of matrices
satisfying the Clifford algebra anticommutation relations (\ref{LD}) in
$N\!-\!1$ dimensions.
As observed above,
the
matrix
\beq
{\bfh}\equiv \tilde{\G}^1\tilde{\G}^2\cdots \tilde{\G}^{N-1}
\eeq
anticommutes with each of the $\tilde{\G}^j$'s. Thus if we define
\beq
\G^N=\e \bfh ,
\eeq
where the constant $\e$ is such that $(\G^N)^2={\bf 1}$, then the matrices
\beq
\G^a=(\tilde{\G}^j,\G^N)
\eeq
satisfy the Clifford algebra (\ref{ACR}) in $N$ dimensions.
It can be shown by induction
that the choice $\epsilon = (-i)^{(N-1)/2}$ gives
\beq
\label{GAMN}
\Gamma^N = \left( \begin{array}{cc} {\bf 1} & {\bf 0} \\
                                    {\bf 0} & {\bf -1}  \end{array} \right)
\eeq
and, as a result,
\beq
\label{GAMALL}
\Gamma^1\Gamma^2\cdots\Gamma^N = i^{(N-1)/2}.
\eeq

Again the matrices $\S^{ab}$ defined by (\ref{GENE}) satisfy (\ref{SONCR}).
This time they generate an irreducible representation of $Spin(N)$,
namely the unique fundamental spinor representation with highest weight
\beq
\label{HW2}
\tau=(\frac{1}{2},\frac{1}{2},\ldots ,\frac{1}{2})
\eeq
(see \cite{BARUT} p.224).
If $\s_+$ and $\s_-$ denote the two fundamental spinor
representations of $Spin(N-1)$ (see Case 1 above), we have the branching rule
\beq
\label{CORPO}
\tau|_{Spin(N-1)}=\s_+\oplus \s_- .
\eeq
Clearly the generators of $\tau|_{Spin(N-1)}$ are the matrices
$\tilde{\S}^{jk}\equiv \frac{1}{4}[\tilde{\G}^j,\tilde{\G}^k]$.
\newpage

\setcounter{equation}{0}
\section{Spherical modes of the Dirac operator}

\subsection{$S^N$}

The metric on $S^N$ may be written in geodesic polar coordinates as
\beq
\label{METRIC}
ds_N^2=d\t^2+f^2(\t)ds^2_{N\!-\!1}=d\t^2+f^2(\t)\sum_{i,j=1}^{N-1}
\tilde{g}_{ij}d\o^i\otimes d\o^j,
\eeq
where $\t$ is the geodesic distance from the origin (north-pole),
$f(\t)=\sin\t$, and $\{\o^i\}$ are coordinates on $S^{N-1}$, with metric
tensor $\tilde{g}_{ij}(\o)=<\pa/\pa\o^i,\pa/\pa\o^j>$. Let
$\{\tilde{\bf e}_j\}$ be a vielbein (i.e. an orthonormal frame)
on $S^{N-1}$,
with
anolonomy and (Levi-Civita) connection coefficients
\beq
[\tilde{\bf e}_i,\tilde{\bf e}_j]=\sum_{k=1}^{N-1}
\tilde{C}_{ijk}\tilde{\bf e}_k,
\eeq
\beq
\tilde{\o}_{ijk}=
<\tilde{\nabla}_{\tilde{\bf e}_i}\tilde{\bf e}_j,\tilde{\bf e}_k>=
\frac{1}{2}(\tilde{C}_{ijk}-\tilde{C}_{ikj}-\tilde{C
}_{jki}).
\eeq

Let $\{\tilde{\bf e}^j\}$ be the dual coframe to $\{\tilde{\bf e}_j\}$.
Then
$ds^2_{N\!-\!1}=\sum_{i=1}^{N-1}
\tilde{\bf
e}^i\otimes \tilde{\bf e}^i$.
We shall work in the geodesic polar coordinates vielbein
$\{{\bf e}_a\}_{a=1,\ldots ,N}$ on $S^N$ defined
by
\beq
\label{VIEL}
{\bf e}_N=\pa_{\t}\equiv \pa /\pa \t,\;\;\;\;\;\;
{\bf e}_j={1\over f(\t)}\tilde{\bf e}_j,\;\;\;\;\;\;j=1,\ldots ,N-1.
\eeq
The only nonvanishing components of the Levi-Civita connection
$\o_{abc}$ in the frame $\{{\bf e}_a\}$ are
found to be
\beq
\label{LCC}
\o_{ijk}=\frac{1}{f}\tilde{\o}_{ijk},\;\;\;\;\;\;\o_{iN
k}=-\o_{ikN}=\frac{f'}{f}\d_{ik},\;\;\;\;i,j,k=1,\ldots ,N-1,
\eeq
where a prime denotes differentiation with respect to the argument.
Note that
$\o_{abc}=\o_{a[bc]}\equiv \frac{1}{2}(\o_{abc}-\o_{acb})$,
as required in a vielbein for a metric
connection.

Spinors are $2^{[\frac{N}{2}]}$-dimensional and
are associated with orthonormal frames.
Under a local rotation $\L:S^N\rightarrow SO(N)$, which transforms
the covielbein as
\beq
{\bf e}^a(x)\rightarrow {\bf e}^{\prime a}(x)=\sum_b\L(x)^a\,_b{\bf e}^b(x),
\eeq
a
spinor
transforms by definition
according to
\beq
\psi(x)\rightarrow \psi '(x)=S(\L(x))\psi(x) ,
\eeq
where
$S(\L(x))$ is either one of the two elements of $Spin(N)$ (more precisely
of the spin representation $\tau$ of $Spin(N)$, see section 2) which
correspond to $\L(x)\in SO(N)$, and is
determined by
\beq
S(\L(x))^{-1}\G^aS(\L(x))=\sum_b\L(x)^a\,_b\G^b.
\eeq

A spin connection on $S^N$ is induced by the Levi-Civita connection $\bfo$.
The covariant derivative $\nabla_{a}\psi$ of a spinor $\psi$
along the vielbein ${\bf e}_a$ may be written
 as
\beq
\label{CDS}
\nabla_{a}\psi={\bf e}_{a}\psi -\frac{1}{2}\o_{abc}\S^{bc}\psi,
\eeq
where
summation over repeated indices is understood from now on.
The Dirac operator is defined by
$\slash{\! \nabla}\psi   = \G^a\nabla_a\psi $. We shall now solve for the
eigenfunctions of $\slash{\! \nabla}$ on $S^N$ by separating variables in
geodesic polar coordinates.

\

Case 1. $N$ even. Using eqs. (\ref{PRIMA}), (\ref{LD}), (\ref{PRIMA2}),
(\ref{LG}), (\ref{LG2}), and (\ref{LCC}),
it is straightforward to derive the following
expression for the Dirac operator in the vielbein (\ref{VIEL}):
\begin{eqnarray}
\slash{\! \nabla}\psi=
 (\pa_{\t}+\frac{N-1}{2}\frac{f'}{f})\G^N \psi+\frac{1}{f}\G^i
(\tilde{\bf
e}_i-\frac{1}{2}\tilde{\o}_{ijk}\S^{jk})\psi \nonumber \\
\label{PACHIE1}
=(\pa_{\t}+\frac{N-1}{2}\frac{f'}{f})\G^N
\psi+\frac{1}{f}
\left( \begin{array}{cc}0&i\,\tilde{\slash{\! \nabla}}
\\-i\,\tilde{\slash{\!
\nabla}}&0 \end{array} \right)\psi ,
\end{eqnarray}
where $i=\sqrt{-1}$ and
\beq
\label{DOLD}
\tilde{\slash{\! \nabla}}=\tilde{\G}^j\tilde{\nabla}_j=
\tilde{\G}^j
(\tilde{\bf
e}_j-\frac{1}{2}\tilde{\o}_{jkl}\tilde{\S}^{kl})
\eeq
is the
Dirac
operator on $S^{N-1}$. We now project the first order Dirac equation
\beq
\label{FODE}
\slash{\! \nabla}\psi =i\l \psi
\eeq
onto the ``upper" and ``lower" components of $\psi$. Define
\beq
\psi\equiv \left( \begin{array}{c}\phi_+ \\\phi_- \end{array}
\right).
\eeq
Then
\beq
\label{DE1}
\left\{ \begin{array}{ll}
(\pa_{\t}+\frac{N-1}{2}\frac{f'}{f})\phi_-
+\frac{1}{f}i\,\tilde{\slash{\!
\nabla}}\phi_-= i\l \phi_+ ,\\
(\pa_{\t}+\frac{N-1}{2}\frac{f'}{f})\phi_+
-\frac{1}{f}i\,\tilde{\slash{\!
\nabla}}\phi_+=
i\l \phi_-.
\end{array}
\right.
\eeq
Eliminating $\phi_-$ (or $\phi_+$) gives the second order equation
\beq
\label{DE0}
\left[(\pa_{\t}+\frac{N-1}{2}\frac{f'}{f})^2+{1\over
f^2}\tilde{\slash{\!
\nabla}}^2 \pm  {f'\over f^2}i\,\tilde{\slash{\!
\nabla}}\right]\phi_{\pm}=-\l^2 \phi_{\pm},
\eeq
which is equivalent to the equation obtained by squaring the Dirac
operator,
i.e.
\beq
\label{IDE}
\slash{\! \nabla}^2\psi=-\l^2 \psi.
\eeq

It is well known that on a compact spin manifold $\slash{\! \nabla}^2$ is
negative semidefinite (the spectrum of $\slash{\! \nabla}$ is purely
imaginary), so that $\l$ is real.
Now suppose that we have solved the eigenvalue equation on $S^{N-1}$,
i.e.
\beq
\label{RECUR}
\tilde{\slash{\! \nabla}}\chi_{lm}^{(\pm)}=\pm
i(l+\r)\chi_{lm}^{(\pm)}.
\eeq
Here,
the index $l=0,1,\ldots ,$ labels the
eigenvalues
of the Dirac operator on the $(N-1)$-sphere,
and $\rho\equiv (N-1)/2$,~\footnote{The
spectrum of the Dirac operator on spheres is
well known (see e.g. \cite{TRAUTMAN}). It will be clear that our procedure
gives an independent proof of (\ref{EIEI})
by induction over the dimension $N$ (see also
the remark at the end of this subsection). Eq. (\ref{RECUR}) (and the
analogous relation (\ref{RECUR2})) may then be assumed as the inductive
hypothesis in this proof.
A group-theoretic derivation of the spectrum of $\slash{\! \nabla}^2$
on $S^N$ will be given in subsections 5.1 and 5.2.}
while the index $m$ runs
from $1$
to the degeneracy $d_l$.  Since the dimension of $\chi$ is the same
as the
dimension of $\phi_{\pm}$, i.e. $2^{N/2-1}$, one can
separate
variables in the following way:
\begin{eqnarray}
\label{SOV1}
^{(1)}\phi_{+nlm}(\t,\O)=\phi_{nl}(\t)\chi_{lm}^{(-)}(\O),\\
\label{SOV2}
^{(2)}\phi_{+nlm}(\t,\O)=\psi_{nl}(\t)\chi_{lm}^{(+)}(\O),
\end{eqnarray}
and similarly for the ``lower" spinor $\phi_-$. Here $\O\in S^{N-1}$, and
$n=0,1,\cdots $ labels the eigenvalues $-\l_{n,N}^2$ of $\slash{\!
\nabla}^2$ on
$S^N$ and $n\geq l$ as is well known. (We shall see also that this condition
arises as the requirement for the
absence of singularity in the mode functions.)

Substituting (\ref{SOV1}) in
(\ref{DE0}) we
obtain the following equation for the scalar functions $\phi_{nl}$:
\beq
\label{EE1}
D\,\phi_{nl}=-\l^2_{n,N}\,\phi_{nl},
\eeq
where $D$ is the differential operator
\begin{eqnarray}
D & = & \left({\pa\over \pa \t}+\r \cot\t\right)^2-{(l+\r)^2\over
\sin^2\t}+(l+\r){\cos\t\over \sin^2\t}\nonumber \\
\label{EE12}
 & = & {\pa^2\over \pa\t^2}+(N-1)\cot\t{\pa\over \pa\t}-
{(l\!+\!\r)^2-
\r(\r\!-\!1)\over \sin^2\t}+
(l\!+\!\r){\cos\t\over \sin^2\t}-\r^2.
\end{eqnarray}
A simple calculation allows us to rewrite $D$ in the following form
\beq
D=(\cos\frac{\t}{2})^{l+1}(\sin\frac{\t}{2})^l\left(D_{\cos\t}^{(\frac{N
}{2}+l-1,
\frac{N}{2}+l)}-(l+\frac{N}{2})^2\right)\circ
(\cos\frac{\t}{2})^{-l-1}(\sin\frac{\t}{2})^{-l},
\eeq
where $D_x^{(a,b)}$ is the differential operator for the Jacobi
polynomials
$P_n^{(a,b)}(x)$ \cite{GRR},
\begin{eqnarray}
D_x^{(a,b)} & = & (1-x^2)\pa_x^2+[b-a-(a+b+2)x]\pa_x ,\\
D_x^{(a,b)} P_n^{(a,b)}(x) & = & -n(n+a+b+1)P_n^{(a,b)}(x).
\end{eqnarray}
Thus, the unique regular solution to (\ref{EE1}) is, up to a
normalization
factor,
\beq
\label{MODES1}
\phi_{nl}(\t)=(\cos\frac{\t}{2})^{l+1}
(\sin\frac{\t}{2})^lP_{n-l}^{(\frac{N}{2}+l
-1,\frac{N}{2}+l)}(\cos\t),
\eeq
with $n-l\geq 0$ -- this condition is needed for the regularity of
the eigenfunctions
-- and with the eigenvalues
\beq
\label{EIEI}
\l_{n,N}^2=(n+\frac{N}{2})^2.
\eeq
At $\t=0$ only the modes with $l=0$ are nonzero. The functions
$\phi_{n0}$ are
called {\em spinor spherical functions}.

By proceeding in a similar way with the functions $\psi_{nl}$ in
(\ref{SOV2})
we find
\begin{eqnarray}
\label{MODES2}
\psi_{nl}(\t) & = & (\cos\frac{\t}{2})^l(\sin\frac{\t}{2})^{l+1}P_{n-l}
^{(\frac{N}{2}+l,\frac{N}{2}+l-1)}(\cos\t) \\
 & = & (-1)^{n-l}\phi_{nl}(\pi-\t).  \label{MODES3}
\end{eqnarray}
They all vanish at the north pole, but for $l=0$ they are nonzero at
the south
pole ($\t=\pi$).

One can readily verify that if $\phi_{+}$ satisfies (\ref{DE0}) and if
$\phi_{-}$ is {\em defined} by the second equation of (\ref{DE1}), then
the first equation is automatically satisfied.  In this way we can find the
eigenfunctions of the first-order Dirac operator.
We use the formulae
\begin{eqnarray}
\left[ \frac{d\ }{d\theta} + \frac{N-1}{2}\cot\theta -
\frac{l+\frac{N-1}{2}}{\sin\theta}\right]\phi_{nl}(\theta) & = &
- (n+\frac{N}{2})\psi_{nl}(\theta),  \label{PHIPSI}\\
\left[ \frac{d\ }{d\theta} + \frac{N-1}{2}\cot\theta +
\frac{l+\frac{N-1}{2}}{\sin\theta}\right]\psi_{nl}(\theta) & = &
 (n+\frac{N}{2})\phi_{nl}(\theta),  \label{PSIPHI}
\end{eqnarray}
to simplify the expressions.  These can be proved by using
the expression of the Jacobi polynomial in terms of the hypergeometric
function,
\beq
\label{JACOBI}
P_{n}^{(\alpha,\beta)}(x) = \frac{\Gamma(n+\alpha+1)}
{\Gamma(n+1)\Gamma(\alpha+1)}F(n+\alpha+\beta+1,-n,\alpha+1;\frac{1-x}{2}),
\end{equation}
and the raising and lowering operators for the third entry of the
hypergeometric function.
Thus, we find the solutions to the first-order Dirac equation to be
\begin{eqnarray}
\psi_{\pm nlm}^{(-)} (\theta,\Omega) & = & {c_N(nl)\over \sqrt{2}}
\left( \begin{array}{c}
\phi_{nl}(\theta)
\chi_{lm}^{(-)}(\Omega) \\ \pm i\psi_{nl}(\theta)\chi_{lm}^{(-)}(\Omega)
\end{array}\right), \label{SOL1} \\
\psi_{\pm nlm}^{(+)} (\theta,\Omega) & = & {c_N(nl)\over \sqrt{2}}
\left( \begin{array}{c}
i \psi_{nl}(\theta)
\chi_{lm}^{(+)}(\Omega) \\ \pm \phi_{nl}(\theta)\chi_{lm}^{(+)}(\Omega)
\end{array}\right).  \label{SOL2}
\end{eqnarray}
These spinors satisfy
\begin{equation}
\label{PACHIE3}
\slash{\! \nabla}\psi_{\pm nlm}^{(s)} = \pm i\left( n + \frac{N}{2}\right)
\psi_{\pm nlm}^{(s)},
\end{equation}
with $s = \pm$, and are required to satisfy the normalization condition
\beq
\label{NOORMAA}
\int_{S^N}d\O_N\psi_{+ nlm}^{(s)} (\theta,\Omega)^{\dagger}
\psi_{+ n'l'm'}^{(s')} (\theta,\Omega)=\d_{nn'}\d_{ll'}\d_{mm'}\d_{ss'},
\eeq
with an analogous relation for $\psi^{(s)}_{+}\rightarrow \psi^{(s)}_-$.

Suppose that the spinors $\chi_{lm}^{(\pm)}(\Omega)$ are normalized by
\beq
\int_{S^{N-1}} d\Omega_{N-1}\chi_{lm}^{(s)}(\Omega)^{\dagger}\chi_{l'm'}^{(s')}
(\Omega) = \delta_{ll'}\delta_{mm'}\d_{ss'}.
\eeq
Then the normalization factor $c_N(nl)$ is determined by
\beq
|c_{N}(nl)|^{-2} = \frac{1}{2}
\int_0^{\pi} d\theta\, \sin^{N-1}\theta\,( \phi_{nl}(\theta)^2 +
\psi_{nl}(\theta)^2 ).
\eeq
It is clear from
(\ref{MODES3})
that the integrals of the first and the
second terms are equal.  Using Ref.\ \cite{GRR} eq. 7.391 n.1 we find
\begin{eqnarray}
|c_{N}(nl)|^{-2} & = &
\int_0^{\pi} d\theta \sin^{N-1}\theta\, \phi_{nl}(\theta)^2 \nonumber \\
 & = & \frac{2^{N-2}\left| \Gamma(\frac{N}{2}+n)\right|^2}
{(n-l)!(N+n+l-1)!}.
\label{NORMALI}
\end{eqnarray}

The spinor eigenfunctions with $l = 0$ near $\theta = 0$ are then given by
\begin{eqnarray}
\psi_{\pm n0m}^{(-)}(\theta,\Omega) & \sim &
\left[{(N+n-1)!\over 2^{N-1}n!|\Gamma(N/2)|^2}\right]^{1/2}
\left(\begin{array}{c}\chi_{0m}^{(-)} \\ 0 \end{array}\right), \\
\psi_{\pm n0m}^{(+)}(\theta,\Omega) & \sim &
\left[{(N+n-1)!\over 2^{N-1}n!|\Gamma(N/2)|^2}\right]^{1/2}
\left(\begin{array}{c} 0 \\ \chi_{0m}^{(+)} \end{array}\right).
\end{eqnarray}
The degeneracy for the eigenvalue $+i(n+N/2)$ [or $-i(n+N/2)$] is given by
(take $n=n'$, $l=l'$, $m=m'$, $s=s'$ in (\ref{NOORMAA}) and sum
over $l,m,s$)
\begin{eqnarray}
D_N(n)  & = &
\int_{S^N} d\Omega_N\,\sum_{slm} \psi_{+nlm}^{(s)}(\theta,\Omega)^{\dagger}
\psi_{+nlm}^{(s)}(\theta,\Omega) \nonumber \\
 & = & \Omega_N \times\lim_{\theta\to 0}\sum_{sm}
\psi_{+n0m}^{(s)}(\theta,\Omega)^{\dagger}
\psi_{+n0m}^{(s)}(\theta,\Omega) \nonumber \\
\label{LULA}
 & = &
{\Omega_N (N+n-1)!\over 2^{N-1}n!|\Gamma(N/2)|^2}
\sum_m \left[ \chi_{0m}^{(+)}(\Omega)^{\dagger}\chi_{0m}^{(+)}(\Omega)
+ \chi_{0m}^{(-)}(\Omega)^{\dagger}\chi_{0m}^{(-)}(\Omega)\right],
\end{eqnarray}
where we have used the fact that the sum over $s,l,m$ inside the integral
is constant over $S^N$ (this is easy to prove),
so that it may be calculated for $\theta\rightarrow
0$, where only the $l=0$ term survives. The factor
$\Omega_N$ is the volume of $S^N$,
\beq
\label{OMEGAN}
\Omega_N = {2\pi^{(N+1)/2}\over \Gamma((N+1)/2)}.
\eeq
 Now the
degeneracy $d_0$ of $+i\r$ (or $-i\r$) on $S^{N-1}$ (cf. (\ref{RECUR}))
coincides with the dimension $2^{N/2-1}$ of the representation of
$Spin(N)$ labelled by $l=0$ (see section 5). Thus
 $m = 1,2,\ldots,2^{N/2 -1}$ in (\ref{LULA}), and we have the identity
\begin{eqnarray}
2^{N/2-1} & = &
\sum_m \int d\Omega_{N-1}\chi_{0m}^{(s)}(\Omega)^{\dagger}\chi_{0m}^{(s)}
(\Omega)  \nonumber \\
 & = & \Omega_{N-1}\sum_m
\chi_{0m}^{(s)}(\Omega)^{\dagger}\chi_{0m}^{(s)}(\Omega).
\end{eqnarray}
{}From this equation and eqs.\ (\ref{LULA}) and (\ref{OMEGAN}) we obtain
\begin{equation}
\label{DIME1}
D_N(n)
  = {2^{N/2}(N+n-1)!\over n!(N-1)!},
\end{equation}
where we have used the doubling formula for the Gamma function
\beq
\label{DFGF}
\Gamma(x)\Gamma(x+\frac{1}{2}) = {\sqrt{\pi}\over 2^{2x-1}}\Gamma(2x).
\eeq
The degeneracy
$D_N(n)$ is equal to the dimension of the spinor representation of
$Spin(N+1)$ labelled by $n$ (see section 5).
Of course, the degeneracy of the eigenvalue $-(n+N/2)^2$ of $\slash{\!
\nabla}^2$ is $2D_N(n)$.

\

Case 2. $N$ odd ($\geq 3$). In this case a Dirac spinor on $S^N$ is
irreducible under
$Spin(N)$ and the dimension of the $\G$-matrices is $2^{(N-1)/2}$,
the same as on
$S^{N-1}$.
The Dirac operator in the geodesic polar coordinates vielbein
(\ref{VIEL})
takes the form
\beq
\label{PACHIE2}
\slash{\! \nabla}\psi=(\pa_{\t}+\r \cot \t)\G^N
\psi+\frac{1}{\sin\t}\tilde{\slash{\! \nabla}}\psi,
\eeq
where $\r=(N-1)/2$ and
$\tilde{\slash{\! \nabla}}=\tilde{\G}^j\tilde{\nabla}_j$ is the
Dirac
operator (\ref{DOLD}) on $S^{N-1}$. Since
\beq
\label{SMART1}
\G^N\tilde{\slash{\! \nabla}}+\tilde{\slash{\! \nabla}}\G^N=0,
\eeq
the iterated Dirac operator is given by
\beq
\label{IDONODD}
\slash{\! \nabla}^2 \psi=(\pa_{\t}+\r \cot\t)^2\psi +\frac{1}{\sin^2\t}
\tilde{\slash{\! \nabla}}^2\psi
-\frac{\cos\t}{\sin^2\t}\G^N\tilde{\slash{\!
\nabla}}\psi .
\eeq

In order to separate variables in the eigenvalue equation (\ref{IDE})
we observe that although $\tilde{\slash{\! \nabla}}$ does not commute
with
$\slash{\! \nabla}^2$, the operator $\G^N \tilde{\slash{\! \nabla}}$
does, as a
consequence of (\ref{SMART1}).
Thus $\slash{\! \nabla}^2$ and $\G^N\tilde{\slash{\! \nabla}}$ can
have common
eigenfunctions. Furthermore, since $(\tilde{\slash{\!
\nabla}})^{\dagger}=-\tilde{\slash{\! \nabla}}$ and
$(\G^N)^{\dagger}=\G^N$,
the operator $\G^N\tilde{\slash{\! \nabla}}$ is hermitian (because of
(\ref{SMART1})), and has real eigenvalues.

Suppose that we have solved the equation
\beq
\label{EENODD}
\G^N\tilde{\slash{\! \nabla}}\hat{\chi}^{(\pm)}_{lm}=\pm k
\hat{\chi}^{(\pm)}_{lm}
\eeq
for the spinor $\hat{\chi}$ on $S^{N-1}$. In order to find the possible
values of $k$ we
apply $\tilde{\slash{\! \nabla}}$ to both sides of (\ref{EENODD}) and
use
(\ref{SMART1}) to find
\beq
\tilde{\slash{\! \nabla}}^2\hat{\chi}_{lm}^{(\pm)}
=-k^2\hat{\chi}_{lm}^{(\pm)}.
\eeq
Thus $-k^2$ are the eigenvalues of the iterated Dirac operator on
$S^{N-1}$,
i.e.
\beq
\label{RECUR2}
k=l+\r,\;\;\;\;\;l=0,1,\ldots .
\eeq
Let $\chi_{lm}^{(-)}$  satisfy
\beq
\tilde{\slash{\! \nabla}}
\chi_{lm}^{(-)} = -ik\chi_{lm}^{(-)}.
\eeq
Then $\chi_{lm}^{(+)} \equiv \Gamma^N\chi_{lm}^{(-)}$  is
the eigenfunction of $\tilde{\slash{\! \nabla}}$
with eigenvalue $+ik$.  These are related to
$\hat{\chi}_{lm}^{(\pm)}$ by
\begin{eqnarray}
\hat{\chi}_{lm}^{(-)} & = &
\frac{1}{\sqrt{2}}(1 + i\Gamma^N)\chi_{lm}^{(-)}, \\
\hat{\chi}_{lm}^{(+)} & = & \Gamma^N\hat{\chi}_{lm}^{(-)}.
\end{eqnarray}

We can now separate variables in (\ref{IDE}) by letting
\begin{eqnarray}
\label{SOV3}
^{(1)}\psi_{nlm}(\t,\O)=\phi_{nl}(\t)\hat{\chi}^{(-)}_{lm}(\O),\\
\label{SOV33}
^{(2)}\psi_{nlm}(\t,\O)=\psi_{nl}(\t)\hat{\chi}^{(+)}_{lm}(\O).
\end{eqnarray}
Substituting (\ref{SOV3}) in (\ref{IDE}) and using (\ref{IDONODD}) we
find that
the scalar functions $\phi_{nl}$ satisfy the same equation
(\ref{EE1}) as in the case of $N$ even. We immediately
conclude
that $\phi_{nl}$ is given by (\ref{MODES1}), and similarly we find
that
$\psi_{nl}$ is given by (\ref{MODES2}).

The eigenfunctions of the first-order Dirac operator must be of the form
\beq
\psi (\t,\O)=
\phi_{nl}(\theta)\hat{\chi}_{lm}^{(-)}(\Omega) + c\psi_{nl}(\theta)
\hat{\chi}_{lm}^{(+)}(\Omega).  \label{SOL3}
\eeq
Then, using (\ref{PHIPSI}) and (\ref{PSIPHI}) in (\ref{PACHIE2}),
we find that $\psi$ is an eigenfunction
of the first-order Dirac operator with eigenvalues $\pm i (n+\frac{N}{2})$
if we choose $c = \pm i$.  The normalization factors are given again by
$c_N(nl)/\sqrt{2}$.  The formula for the dimensionality is the same as that for
$N$ even except that the $l=0$ representation of $Spin(N)$
is $2^{\frac{N-1}{2}}$-dimensional (and there is no factor of $2$ we had for
$N$ even).  Thus, the dimensionality is
\beq
D_N(n) =
{2^{(N-1)/2}(N+n-1)!\over n!(N-1)!}.
\eeq

Our method shows how to separate variables in the Dirac equation
written
in geodesic polar coordinates. The spinor modes of $\slash{\!
\nabla}^2$ on
$S^N$ are the product of the spherical modes $\phi_{nl}$, $\psi_{nl}$
times the
spinor modes on $S^{N-1}$ [(\ref{RECUR}) ($N$ even) or (\ref{EENODD})
($N$ odd)].
But again these can be obtained
from the
spinor modes on $S^{N-2}$, and so on.  Therefore we have set up an
induction
procedure by which the spinor modes on $S^N$ can be recursively
calculated
starting from the spinor modes on the lowest dimensional sphere
$S^2$.

Note also that our procedure gives an independent proof of the spectrum of the
Dirac operator $\slash{\! \nabla}$ by induction.
First consider the case $N = 2$.  Writing
$ds_2^2 = d\theta^2 + \sin^2\theta d\varphi^2$, one simply has
$\tilde{\slash{\! \nabla}} = \partial/\partial\varphi$.
We need to find which eigenvalues of this operator are allowed.  Spinors on
$S^2$ should transform as a double-valued spin-1/2 representation of $SO(2)$
under the rotation of zweibein.  Now, the loop defined by
$0 \leq \varphi < 2\pi$ with $\theta =$ const. in the bundle of frames over
$S^2$ is homotope to the $2\pi$-rotation at a point for our system of
zweibein.  Hence, the spinor field must change sign when it goes around this
loop.  Thus, we have
$\partial/\partial\varphi = \pm i/2,\pm 3i/2,\ldots$~\footnote{It also
follows that
the spinor must vanish both at $\theta=0$ and $\pi$ if
$\partial/\partial\varphi \neq \pm i/2$.  This is automatically satisfied by
requiring regularity of the solutions.}
The modes are given by letting
$\chi_{lm}^{(\pm)}(\Omega)  = e^{\pm i(l + 1/2)\varphi}$ ($l=0,1,\ldots $)
in (\ref{SOV1}) and
(\ref{SOV2}).
Then regular solutions of the Dirac equation
$\slash{\! \nabla}\psi = i\lambda\psi$ on $S^2$
exist only for $\lambda = \pm (n+1)$
with $n (\geq l)$ being a nonnegative integer.  Now, assume that the spectrum
of the operator $\tilde{\slash{\! \nabla}}$ on $S^{N-1}$ is $\pm
i(l+\rho)$, $\r=(N-1)/2$, $l=0,1,\ldots$
Then, whether $N$ is even or odd, regular solutions of the Dirac equation on
$S^{N}$ exist only for $\lambda = \pm (n + N/2)$ with $n (\geq l)$ being a
nonnegative integer.  This completes the proof by induction that the spectrum
of the Dirac operator on $S^N$ is indeed given by $\pm i(n + N/2)$ with
$n=0,1,\ldots$

\subsection{$H^N$}

The real
hyperbolic space $H^N$ is the noncompact partner of $S^N$. The metric
on $H^N$ in
geodesic polar coordinates takes the form (\ref{METRIC}) with
$\t\rightarrow y$
and $f(y)=\sinh y$, where $y$ is the geodesic distance from the
origin. By
repeating the same steps as on $S^N$ one obtains a hypergeometric
equation for
the spherical modes $\phi_{\l l}(y)$. The spectrum of $\slash{\!
\nabla}^2$ is
now continuous and is labelled by the real parameter $\l$ (as in
(\ref{IDE})).
The final results on $H^N$ and $S^N$ are related by analytic
continuation in
the geodesic distance. More precisely, one expresses the Jacobi
polynomials in
terms of the hypergeometric function according to (\ref{JACOBI})
and then one makes the replacements
\beq
\t\rightarrow iy,\;\;\;\;\;\;n\rightarrow -i\l-\frac{N}{2}
\eeq
in the spinor modes found above on $S^N$.
The result for the
(unnormalized)
spherical modes on $H^N$ is
\beq
\label{DODO1}
\phi_{\l
l}(y)=(\cosh\frac{y}{2})^{l+1}(\sinh\frac{y}{2})^lF(\frac{N}{2}
\!+\!l\!+\!i\l,\frac{N}{2}\!+\!l\!-\!i\l,\frac{N}{2}\!+\!l,-\sinh^2
\frac{y}{2}).
\eeq
For the modes $\psi_{\l l}$ we find similarly
\beq
\label{DODO2}
\psi_{\l
l}(y)=\frac{2\l}{N+2l}
(\cosh\frac{y}{2})^{l}(\sinh\frac{y}{2})^{l+1}F(\frac{N}{2}
\!+\!l\!+\!i\l,\frac{N}{2}\!+\!l\!-\!i\l,\frac{N}{2}\!+\!l\!+\!1,
-\sinh^2\frac{y}{2}).
\eeq
Here $\l\in {\bf R}$, and $l=0,1,\ldots ,\infty$.
The eigenfunctions of the first-order Dirac operator can also be obtained by
letting $\theta \to iy$  and $n\rightarrow -i\l-\frac{N}{2}$ in
(\ref{SOL1})-(\ref{SOL2}) ($N$ even) and in (\ref{SOL3}) ($N$ odd). Notice
from (\ref{PACHIE1}) and (\ref{PACHIE2}) that
$\slash{\! \nabla}|_{S^N}\rightarrow \frac{1}{i}\slash{\! \nabla}|_{H^N}$
for $\theta \to iy$.
It follows
from (\ref{PACHIE3})
that the solution of the Dirac equation on $H^N$ with
eigenvalue $i\l$ is obtained from that on $S^N$ with eigenvalue
$i(n+N/2)$.
Thus we find for $N$ even
\beq
\label{PACHIE4}
\slash{\! \nabla}\psi^{(-)}_{\pm\lambda l m}
= \pm i\lambda \psi^{(-)}_{\pm\lambda l m},
\eeq
where
\beq
\label{NCSSM}
\psi^{(-)}_{\pm\lambda l m}(y,\O) \equiv {c_{N}(\lambda l)\over \sqrt{2}}
\left( \begin{array}{c} \phi_{\lambda l}(y)\chi_{lm}^{(-)}(\Omega) \\ \pm i
\psi_{\l l}(y)\chi_{lm}^{(-)}(\Omega)\end{array}\right),
\eeq
where $c_{N}(\l l)$ is a normalization constant.
The modes $\psi^{(+)}_{\pm\lambda l m}$ are obtained by interchanging
$\phi_{\l l}(y)$ and $\psi_{\l l}(y)$, and letting
$\chi_{lm}^{(-)} \to \chi_{lm}^{(+)}$ on the right-hand side.
The asymptotic behavior for large $y$ is obtained as in the scalar case
(see, e.g. \cite{CAMPOHIGU})
as
\begin{eqnarray}
\phi_{\l l}(y) & \sim & c_{l}(\lambda)e^{(-\rho + i\lambda)y} + {\rm c.c.} \\
\psi_{\l l}(y) & \sim & i\left[
c_{l}(\lambda)e^{(-\rho + i\lambda)y} - {\rm c.c.}\right],
\end{eqnarray}
where
\beq
c_{l}(\lambda) = {2^{N-2}\over \sqrt{\pi}}
{\Gamma(N/2+l)\Gamma(1/2+i\lambda)\over
\Gamma(N/2+l+i\lambda)}.
\eeq
Then the (continuous-spectrum)
normalization constant can be found, again, with the method used
for scalar and tensor fields \cite{CAMPOHIGU} as
\beq
\label{DODO3}
|c_{N}(\lambda l)|^{2} = c_N |c_{l}(\l )|^{-2},
\eeq
where $c_N=2^{N-2}/\pi$.
We define the spectral function $\m (\l)$, which is needed to find the
heat kernel, by
\beq
\mu(\l) \equiv {\Omega_{N-1}\over c_N g(1/2)}
\sum_{slm}\psi^{(s)}_{+\l lm}(0)^{\dagger}\psi^{(s)}_{+\l lm}(0),
\eeq
where again $s=\pm$ and
the spin factor $g(\frac{1}{2}) = 2^{N/2}$ is the dimension of the
spinor.  Then we find
\beq
\label{SSF}
\m(\l) = |c_{0}(\lambda)|^{-2}
= {\pi\over 2^{2N-4}}\left| {\Gamma(N/2+i\l)\over
\Gamma(N/2)\Gamma(1/2+i\l)}\right|^2.
\eeq
The computation proceeds almost in the same way for $N$ odd
and we obtain the same
formula for $\m(\l)$.
More explicitly the spectral function is given by
\beq
\m(\l)={\pi\over 2^{2N-4}\G(N/2)^2}\left\{ \begin{array}{ll}
\prod_{j=1/2}^
{(N\!-\!2)/2}(\l^2+j^2),\;\;\;\;\;\;N \;odd,\\
 \l\,\coth(\pi \l)
\prod_{j=1}^{(N\!-\!2)/2}(\l^2+j^2),\;\;\;\;\;\;N\;even.
\end{array}
\right.
\eeq

\newpage
\setcounter{equation}{0}
\section{The spinor heat kernel}
\subsection{The compact case}

 The spinor heat kernel $K(x,x',t)$ on $S^N$ satisfies the heat
equation for $\slash{\!\nabla}^2$
\beq
\label{HE}
(-\frac{\pa}{\pa t}+\slash{\!\nabla}_x^2)K(x,x',t)=0,
\eeq
with the initial condition
\beq
\lim_{t\rightarrow 0}\int_{S^N} K(x,x',t)\psi(x')dx'=\psi(x).
\eeq
Here $t$ is the time parameter in the heat equation and we are
suppressing all
spinor indices. $K$ is actually a
$2^{[\frac{N}{2}]}\times
2^{[\frac{N}{2}]}$ matrix. $K(x,x',t)$ maps a spinor at $x'$ to a spinor at
$x$.

\

Case 1. $N$ odd.
Consider the normalized eigenfunctions of $\slash{\! \nabla}$ on $S^N$ (cf.
(\ref{SOL3}))
\beq
\psi^{(\pm)}_{nlm}(\theta,\Omega) = {c_N(nl)\over \sqrt{2}}\left[
\phi_{nl}(\theta)\hat{\chi}_{lm}^{(-)}(\Omega) \pm i\psi_{nl}(\theta)
\hat{\chi}_{lm}^{(+)}(\Omega)\right].  \label{SOLE}
\eeq
Here $n=0,1,\ldots ,\infty$, $l=0,1, \ldots ,n$, and $m=1,\ldots ,
D_{N-1}(l)$.

The heat kernel is given by the
familiar mode expansion
\begin{eqnarray}
\label{HK1}
K(\theta,\O,\theta',\O',t)=\sum_{n,l,m}\left[\psi^{(-)}_{nlm}
(\theta,\Omega) \otimes \psi^{(-)}_{nlm}(\theta',\Omega')^{\ast}
 \right. \nonumber \\
\left. +\psi^{(+)}_{nlm}
(\theta,\Omega) \otimes \psi^{(+)}_{nlm}(\theta',\Omega')^{\ast}\right]
e^{-t(n+N/2)^2}.
\end{eqnarray}
Using (\ref{SOLE}) in (\ref{HK1}) we obtain
\begin{eqnarray}
\label{HK2}
K(\theta,\O,\theta',\O',t)=\sum_{n,l,m}|c_N(nl)|^2 \left[
\phi_{nl}(\theta)\phi_{nl}(\theta')\hat{\chi}_{lm}^{(-)}(\Omega)
\otimes \hat{\chi}_{lm}^{(-)}(\Omega')^{\ast}\right. \nonumber \\
\left.
+\psi_{nl}(\theta)\psi_{nl}(\theta')\hat{\chi}_{lm}^{(+)}(\Omega)
\otimes \hat{\chi}_{lm}^{(+)}(\Omega')^{\ast}\right] e^{-t(n+N/2)^2}.
\end{eqnarray}
The same result would have been obtained by expanding $K$ directly in terms
of the (normalized) eigenfunctions (\ref{SOV3})-(\ref{SOV33}).

In order to simplify the above expression for $K$ we let $\O=\O'$,
i.e., we assume that the two points $x=(\theta,\O)$ and
$x'=(\theta',\O')$ lie on the same ``meridian". There is no loss of
generality in doing so. Then the sum over $m$ can be performed explicitly
by using
\beq
\label{IDEN1}
\sum_m \hat{\chi}_{lm}^{(-)}(\Omega)\otimes
\hat{\chi}_{lm}^{(-)}(\Omega)^{*}
= \sum_m \hat{\chi}_{lm}^{(+)}(\Omega)\otimes
\hat{\chi}_{lm}^{(+)}(\Omega)^{*}
= {D_{N-1}(l)\over D_{N-1}(0)\Omega_{N-1}}{\bf 1}.
\eeq
This equation can be proved by induction in $N$. Thus
\beq
\label{HK3}
K(\theta,\O,\theta',\O,t)={\bf 1}\sum_{n,l}
{D_{N-1}(l)|c_N(nl)|^2\over D_{N-1}(0)\O_{N-1}} \left[
\phi_{nl}(\theta)\phi_{nl}(\theta')+
\psi_{nl}(\theta)\psi_{nl}(\theta')\right] e^{-t(n+N/2)^2}.
\eeq

Notice that the unit matrix ${\bf 1}$ is just the parallel spinor
propagator $U(x,x')$ between $x'$ and $x$. [This is the operator that
parallel transports a spinor from $x'$ to $x$ along the shortest geodesic
(great circle) between them. We assume here that $x$ and $x'$ are not
conjugate.] Indeed the polar frame (\ref{VIEL}) is parallel along the
meridian $\O=\O'=$ constant, and therefore parallel propagation along this
curve is trivial in this frame. We rewrite (\ref{HK3}) as
\beq
K(\theta,\O,\theta',\O,t)=U(x,x')f_N(\theta,\theta',t),
\eeq
where $f_N(\theta,\theta',t)$ denotes the scalar sum over $n,l$
in the right-hand side of (\ref{HK3}).
Notice that any dependence on $\O$ has
dropped out in $f_N$.   Therefore,
 we can let $\theta=0$, i.e, we
can assume that $x$ coincides with the north-pole.
Recalling that $\psi_{nl}(0)=0\;\;\forall n,l$, and $\phi_{nl}(0)=0\;\;
\forall l>0$, we see that only the term with $l=0$ survives in the sum.
Thus we get (writing $\theta$ in place of $\theta '$)
\beq
f_N(\theta,t)\equiv f_N(0,\theta,t)=
\sum_{n=0}^{\infty}
{|c_N(n0)|^2\over \O_{N-1}}
\phi_{n0}(0)\phi_{n0}(\theta)e^{-t(n+N/2)^2}.
\eeq
Using (set $x=1$ in eq. (\ref{JACOBI}))
\beq
P_n^{(\a,\b)}(1)={\G(n+\a+1)\over n!\G(\a+1)},
\eeq
we get from (\ref{MODES1})
\beq
\phi_{n0}(0)={\G(n+N/2)\over n!\G(N/2)}.
\eeq
Recalling (\ref{NORMALI}) and using (\ref{DFGF}) we obtain
\beq
\label{HK4}
f_N(\theta,t)=
\frac{1}{\O_N}\sum_{n=0}^{\infty}d_n\phi_n(\theta)e^{-t(n+N/2)^2},
\eeq
where the $\phi_n(\theta)$ are the functions $\phi_{n0}(\theta)$
normalized by $\phi_n(0)=1$,
\beq
\label{SSF1}
\phi_n(\theta)={\phi_{n0}(\t)\over \phi_{n0}(0)}=
{n!\,\G(\frac{N}{2})\over
\G(n\!+\!\frac{N}{2})}\,cos\frac{\t}{2}\,P_n^{(
\frac{N}{2}-1,\frac{N}{2})}(cos\,\t),
\eeq
and $d_n$ are the degeneracies of
$\slash{\!\nabla}^2$ on $S^N$ without the spin factor
$2^{[\frac{N}{2}]}$:
\beq
\label{SDSNAA}
d_n={2\,(n\!+\!N\!-\!1)!\over n!\,(N\!-\!1)!}.
\eeq
Eqs. (\ref{HK4}), (\ref{SSF1}) and (\ref{SDSNAA}) coincide with eqs.
(3.22), (3.19), and (3.21) of ref. \cite{CAMPO1}.

The spinor heat kernel on $S^N$, $N$ odd, may finally be written as
\beq
\label{HK5}
K(x,x',t)=U(x,x')f_N(d(x,x'),t),
\eeq
where $d(x,x')$ denotes the geodesic distance between $x$ and $x'$.
This form of $K$ is now valid for $x$ and $x'$ arbitrary (nonconjugate)
points on $S^N$.
If $x$ and $x'$ are conjugate (e.g.,
$x$ is the north-pole and $x'$
the south-pole),  the parallel propagator $U(x,x')$ is clearly undefined
(there are infinitely many great semicircles connecting $x'$ to $x$ and $U$
depends on the one we choose to parallel propagate). However for
conjugate points $d(x,x')=\pi$ and the heat kernel (\ref{HK5}) is well
defined and vanishes (since $f_N(\pi,t)=0$).

\

Case 2. $N$ even. Consider the normalized eigenfunctions
of $\slash{\! \nabla}$ on $S^N$ given by (\ref{SOL1}) and
(\ref{SOL2}). The heat kernel of $\slash{\! \nabla}^2$ is given by the mode
expansion
\begin{eqnarray}
K(\t,\O,\t',\O',t)\!=\!\!\sum_{n,l,m}\!\!
\left[\psi_{- nlm}^{(-)} (\theta,\Omega)
\!\otimes \!\psi_{- nlm}^{(-)} (\theta',\Omega')^{\ast}
\!+\!\psi_{+ nlm}^{(-)} (\theta,\Omega)
\!\otimes \!\psi_{+ nlm}^{(-)} (\theta',\Omega')^{\ast}\right. \nonumber \\
\label{OLSO}
\left. \!\!\!\!\!\!+\psi_{- nlm}^{(+)} (\theta,\Omega)\!\otimes \!
\psi_{- nlm}^{(+)} (\theta',\Omega')^{\ast}
\!+\!\psi_{+ nlm}^{(+)} (\theta,\Omega)
\!\otimes \!\psi_{+ nlm}^{(+)} (\theta',\Omega')^{\ast}\right]\!
\!e^{-t(n+N/2)^2}.
\end{eqnarray}
Substituting
(\ref{SOL1})-(\ref{SOL2}) in (\ref{OLSO}) we obtain
\beq
\label{ULTI5}
K(\t,\O,\t',\O',t)=\sum_{n,l,m}|c_N(nl)|^2
\left( \begin{array}{cc}K_{nlm}^+&0 \\  0&K_{nlm}^-
\end{array}\right)e^{-t(n+N/2)^2},
\eeq
where
\begin{eqnarray}
K^+_{nlm}(\t,\O,\t',\O')=
\phi_{nl}(\theta)\phi_{nl}(\theta'){\chi}_{lm}^{(-)}(\Omega)\otimes
{\chi}_{lm}^{(-)}(\Omega')^{\ast}\nonumber \\
+\psi_{nl}(\theta)\psi_{nl}(\theta'){\chi}_{lm}^{(+)}(\Omega)\otimes
{\chi}_{lm}^{(+)}(\Omega')^{\ast},
\end{eqnarray}
and
\begin{eqnarray}
K^-_{nlm}(\t,\O,\t',\O')=
\phi_{nl}(\theta)\phi_{nl}(\theta'){\chi}_{lm}^{(+)}(\Omega)\otimes
{\chi}_{lm}^{(+)}(\Omega')^{\ast}\nonumber \\
+\psi_{nl}(\theta)\psi_{nl}(\theta'){\chi}_{lm}^{(-)}(\Omega)\otimes
{\chi}_{lm}^{(-)}(\Omega')^{\ast}.
\end{eqnarray}
Note that $K=K^+\oplus K^-$, where $K^{\pm}$ are the heat kernels for
``upper" spinors $\psi^+\in \G(E^{\tau_+})$ and ``lower" spinors
$\psi^-\in \G(E^{\tau_-})$, respectively (see section 5 for the notations).

Letting, as before, $\O=\O'$, we can do the sum over $m$
in (\ref{ULTI5}) as eq.
(\ref{IDEN1}) is still valid (with $\hat{\chi}^{(\pm)}\rightarrow
{\chi}^{(\pm)}$). The result is
\beq
K(\t,\O,\t',\O,t)=\left( \begin{array}{cc}{\bf 1}&0 \\  0&{\bf 1}
\end{array}\right)f_N(\t,\t',t),
\eeq
where $f_N(\t,\t',t)$ denotes the sum over $n,l$
in the right-hand side
of (\ref{HK3}).

Again $diag({\bf 1},{\bf 1})=U(x,x')$, the parallel spinor
propagator in the polar frame. Since our manipulations of $f_N(\t,\t',t)$
were independent of $N$ being odd or even, we obtain, by letting
$\t\rightarrow 0$, $f_N(0,\t,t)=f_N(\t,t)$ given by
(\ref{HK4}). Finally the spinor heat kernel on $S^N$, $N$ even, takes the
same form (\ref{HK5}).

\subsection{The noncompact case}

Case 1. $N$ odd. The normalized eigenfunctions of $\slash{\! \nabla}$ on $H^N$
are (cf. (\ref{SOLE}))
\beq
\psi^{(\pm)}_{\l lm}(y,\Omega) = {c_N(\l l)\over \sqrt{2}}\left[
\phi_{\l l}(y)\hat{\chi}_{lm}^{(-)}(\Omega) \pm i\psi_{\l l}(y)
\hat{\chi}_{lm}^{(+)}(\Omega)\right], \label{SOLEMIO}
\eeq
where $\l$ is a real
number and may be taken to be positive,
$l=0,1,\ldots ,\infty$, and $m=1,\ldots ,
D_{N-1}(l)$. The functions $\phi_{\l l}$, $\psi_{\l l}$, and the
normalization factor $c_N(\l l)$ are given by eqs. (\ref{DODO1}),
(\ref{DODO2}), and (\ref{DODO3}), respectively. The
modes (\ref{SOLEMIO})
are eigenfunctions of $\slash{\! \nabla}$ with eigenvalues $\pm i\l$
(cf. (\ref{PACHIE4})). They satisfy the continuous spectrum normalization
\beq
\int_0^{+\infty}\int_{S^{N-1}}\psi^{(s)}_{\l lm}(y,\Omega)^{\dagger}
\psi^{(s')}_{\l'
l'm'}(y,\Omega)(\sinh\,y)^{N-1}dyd\O_{N-1}=\d(\l-\l')\d_{ll'}
\d_{mm'}\d_{ss'},
\eeq
where $s,s'=\pm$ and $\d(\l-\l')$ is the Dirac distribution.
The heat kernel of $\slash{\! \nabla}^2$ is given by
\begin{eqnarray}
\label{HKNC1}
K(y,\O,y',\O',t)=\int_0^{+\infty}\sum_{l,m}\left[\psi^{(-)}_{\l lm}
(y,\Omega) \otimes \psi^{(-)}_{\l lm}(y',\Omega')^{\ast}
+ \right. \nonumber \\
\left. \psi^{(+)}_{\l lm}
(y,\Omega) \otimes \psi^{(+)}_{\l lm}(y',\Omega')^{\ast}\right]
e^{-t\l^2}d\l\\
\label{HKNC12}
=\int_0^{+\infty}\sum_{l,m}|c_N(\l l)|^2 \left[
\phi_{\l l}(y)\phi_{\l l}(y')\hat{\chi}_{lm}^{(-)}(\Omega)
\otimes \hat{\chi}_{lm}^{(-)}(\Omega')^{\ast}+\right. \nonumber \\
\left.
\psi_{\l l}(y)\psi_{\l l}(y')\hat{\chi}_{lm}^{(+)}(\Omega)
\otimes \hat{\chi}_{lm}^{(+)}(\Omega')^{\ast}\right] e^{-t\l^2}.
\end{eqnarray}

We now proceed like in the case of $S^N$. We set $\O=\O'$ and do the sum
over $m$ using (\ref{IDEN1}). Then
we let $y=0$, i.e. we fix the point $x$ at the origin.
The final result, for arbitrary points $x,x'\in H^N$, is
\beq
\label{HK8}
K(x,x',t)=U(x,x')\hat{f}_N(d(x,x'),t),
\eeq
\beq
\label{HK9}
\hat{f}_N(y,t)={2^{N-3}\G(N/2)\over
\pi^{N/2+1}}\int_0^{+\infty}\phi_{\l}(y) e^{-t\l^2}\m(\l)d\l,
\eeq
where $\phi_{\l}=\phi_{\l0}$, $\m(\l)$ is the spinor spectral function
(\ref{SSF}), and $U$ is the parallel propagator.
 This agrees with eq. (5.14) of ref. \cite{CAMPO1}.

\

Case 2. $N$ even. The normalized eigenfunctions of $\slash{\! \nabla}$
are given by (\ref{NCSSM}). In terms of these modes the heat kernel is
given by a formula similar to (\ref{OLSO}), with $\sum_n\rightarrow
\int_0^{\infty}d\l$. The computation proceeds in the same way as for $N$
odd
and we obtain for $K$ the same expression (\ref{HK8})-(\ref{HK9}), again in
agreement with ref. \cite{CAMPO1}.

\newpage
\setcounter{equation}{0}
\section{Group theoretic derivation of the heat kernel}

In this section we give a group-theoretic derivation of the spectrum, the
eigenfunctions, and the heat kernel of the iterated Dirac operator on $S^N$
and $H^N$.  We
begin by reviewing some basic facts about harmonic analysis
for homogeneous
vector bundles over compact symmetric spaces (SS) (see, e.g. \cite{CAMPO2}
appendix B).

\subsection{Harmonic analysis on compact symmetric spaces}

Let $U$ be a compact semisimple Lie group and
$K$ a closed subgroup such that the
coset space $U/K$ is diffeomorphic to a Riemannian symmetric space of the
compact type. Let $x\rightarrow ux$ ($u\in U$, $x\in U/K$) be
the natural left-action of $U$ on $U/K$, and let $\pi$ denote
the projection of $U$ onto $U/K$,
$\pi(u)=uK$.

The coset $x_0=eK$ is called the {\em origin} of $U/K$.
Let $\s :{\cal O}\subset
U/K\rightarrow U$ be a local section of the principal bundle
$U(U/K,K)$ (see e.g. ref. \cite{KN}), i.e. a smooth assignment of an
element
$\s(x)\in U$ to any $x$ in an open neighborhood ${\cal O}$ of $x_0$
such that $\pi
(\s(x))=x$. Then $\s(x)x_0=x$.
Let ${\cal U}$ and ${\cal K}$ be
the Lie algebras  of $U$ and $K$, respectively, and let
$Ad(U)$ denote the adjoint representation of $U$ in ${\cal U}$.
Since $K$ is compact, $U/K$
is a reductive coset space \cite{NOM}, i.e. we have a direct sum decomposition
${\cal U} = {\cal K} \oplus {\cal P}$, where ${\cal P}$ is a complementary
$Ad(K)$-invariant subspace of ${\cal U}$. ${\cal P}$ may be identified with
the tangent space $T_{x_0}(U/K)$. The natural choice for ${\cal P}$
(that will be made in the following) is
${\cal P}={\cal K}^{\perp}$, the orthogonal complement of ${\cal K}$ in ${\cal
U}$ with respect to the Killing form.
We shall assume that a $U$-invariant metric has been fixed on $U/K$
by restricting the opposite of the Killing form (or a multiple of it) from
${\cal U}$ to ${\cal P}$.

 Let $Exp$ and $exp$ denote
the
exponential mappings on $U/K$ and $U$, respectively. For a symmetric space
it is well known (see \cite{HELGA0}) that $ExpX=\pi(expX)$, $X\in {\cal P}$.
Thus
on a SS a natural
choice for
$\s$ is
\beq
\s(Exp X)=exp X, \;\;\;\;\;\;\;X\in {\cal P}_0,
\eeq
or equivalently $\s(expX\,x_0)=exp X$,
where ${\cal P}_0\subset {\cal P}$ is the largest open
domain of ${\cal P}$ containing the origin and such that the exponential map
(restricted to ${\cal P}_0$) is a
diffeomorphism.

Let $\pa{\cal P}_0$ denote the boundary of ${\cal P}_0$.
Then $\s$ maps $U/K\setminus {\cal C}_0$ onto $exp({\cal P}_0)\subset U$,
where ${\cal C}_0=Exp(\pa{\cal P}_0)$ is called the {\em cut locus} of
$x_0$. It is known that ${\cal C}_0$ consists either of points that are
conjugate to $x_0$, or of points that can be joined to $x_0$ by more than
one minimizing geodesic (see \cite{KN} vol.II p.97 Theorem 7.1). In any
case ${\cal C}_0$ has always zero measure.

The above choice of local section $\s$ will be assumed in the
following. Let $\{Y_i\}$ be an orthonormal basis
of ${\cal P}$. Then the moving frame  $\{Y_i^{\ast}\}$
on $U/K\setminus {\cal C}_0$ defined by
$Y_i^{\ast}(ExpX)=(expX)_{\ast \mid x_0}Y_i$ is orthonormal, and it
is  parallel
along each geodesic emanating from the origin (see \cite{HELGA2} Theorem
3.3 p.208). This is the local frame we shall work with.

Consider a field $x\rightarrow {\bfpsi}(x)$  on $U/K$ which
is a cross section of the homogeneous vector bundle $E^{\tau}$
on $U/K$ defined
by a finite dimensional linear representation $\tau$ of the isotropy
group $K$ (see \cite{WALLACH} section 5.2).
For example scalars, vectors and spinors
correspond to the trivial, isotropy, and spinor representations of
$K$,
respectively.~\footnote{The spinor representation of $K$ is obtained  by
lifting to $Spin(N)$, $N=dimU/K$, the Lie homomorphism of $K$  into $SO(N)$
induced by
the isotropy representation $k\rightarrow k_{\ast \mid x_0}$
by using an orthonormal basis of
$T_{x_0}(U/K)$.
There may be
topological obstructions to the (global) existence of spinors.
 We recall that an oriented Riemannian manifold $M$ admits a spin structure
if and only if the second
Stiefel-Whitney class of $M$ vanishes.}
The vector space of all sections of $E^{\tau}$ will be denoted by
$\G(E^{\tau})$.
We shall assume that the representation  $\tau$ of
$K$ be irreducible of  dimension $d_{\tau}$. The
representation space of $\tau$, denoted $V_{\tau}$, will be identified with
the typical fibre $E_{x_0}$ of $E^{\tau}$ at the origin.

By definition of homogeneous
vector bundle (see \cite{WALLACH} 5.2.1),
the group $U$ acts on $E^{\tau}$ so that
the fiber at $x$ is
mapped by $u$ isomorphically onto the fiber at $ux$, for each $x\in U/K$
and each $u\in U$. Notice that $\s(x)E_{x_0}=E_x$.
The map $expX$ ($X\in {\cal P}$)
between the fibres at $x_0$ and $x=ExpX$ is the
same as the map of parallel transport (in
the vector bundle)  from $x_0$ to $x$ along the geodesic $\g_{
X}(t)=exp(tX)x_0$. In other words, the parallel displacement of
any ${\bf v}\in E_{x_0}$ along $\g_X(t)$ is {\em defined} to be
$exp(tX){\bf v}$. Let $X^{\ast}$ be the vector field
on $U/K\setminus {\cal C}_0$ defined by
$X^{\ast}(ExpY)=(expY)_{\ast \mid x_0}X$ ($X\in {\cal P}$).
It is then easy to
see that the geodesic $\g_{X}(t)=exp(tX)x_0$ is the integral curve through
the origin of $X^{\ast}$, i.e. $d\g_X /dt=X^{\ast}(\g_X(t))$.
The covariant derivative $\nabla_{X^{\ast}}\bfpsi$ of
 $\bfpsi\in \G(E^{\tau})$ at $x=x_0$ is then defined by
\beq
\label{CD34}
(\nabla_{X^{\ast}}\bfpsi)(x_0)=\lim_{t\rightarrow 0}\frac{1}{t}[
exp(-tX)\bfpsi(\g_X(t)) - \bfpsi(x_0)].
\eeq
When $E^{\tau}$ is the tangent bundle on $U/K$, the $U$-invariant
affine connection defined this way is called the {\em canonical
connection of the second kind}, see \cite{NOM}.
The elements $expX$ ($X\in {\cal P}$) are called {\em transvections}, see
\cite{HELGA2} p.209.

Let $\{{\bf v}_a\}$ be a fixed orthonormal basis of $V_{\tau}\simeq
E_{x_0}$.
Then we choose as basis for the fibre $E_x$ at
$x=Exp X$ ($X\in {\cal P}_0$)
the set $\{{\bft}_a(x)\}$ obtained by
transforming $\{{\bf v}_a\}$ with  $\sigma(x)=exp X$,
i.e. ${\bft}_a(x)=\s(x){\bf v}_a$.
 The field $\bfpsi$ is written as
${\bfpsi}(x)=\sum_{a=1}^{d_{\tau}} \psi^a(x){\bft}_a(x)$ for $x\in
U/K\setminus {\cal C}_0$.

The set of
equivalence classes of irreducible unitary representations (irreps)
of $U$ and
$K$ will be
denoted by $\hat{U}$ and $\hat{K}$, respectively.
A simple application of the
Frobenius
Reciprocity Theorem to the induced representation of $U$ on
$L^2(U/K,E^{\tau})$ (the space of square-integrable sections of
$E^{\tau}$)
gives the following harmonic expansion for the components $\psi^a(x)$
(here $\s^{-1}(x)\equiv (\s(x))^{-1}$, see \cite{CAMPO2}
appendix B)
\beq
\label{UNO}
\psi^a(x)=\sum_{\l\in
\hat{U}(\tau)}\sum_{I=1}^{d_{\l}}\sum_{\xi=1}^{\xi_{\l}}
\psi^I_{\l \xi}\, U^{\l}(\s^{-1}(x))^{a\xi}\,_ I,\;\;\;\;\;\;a=1,\ldots
,d_{\tau}.
\eeq
Here $\hat{U}(\tau)$ is the set of irreps $\l\in \hat{U}$ (of dimension
$d_{\l}$)
such that
$\l|_{K}$
contains $\tau$ at least once. The index $\xi$ labels the
multiplicity
$\xi_{\l}$ of $\tau$ in $\l|_{K}$, thus $\xi=1,\ldots ,\xi_{\l}$. Each
representation space $V_{\l}$ is decomposed as $\left( \oplus_{\xi} V_{\xi}
\right) \oplus
V_{rest}$, where  $\l|_{K}$
is
equivalent to $\tau$ in each
$V_{\xi}$, and $V_{rest}=(\oplus_{\xi}V_{\xi})^{\perp}$.
Thus each $V_{\xi}$ is isomorphic to $V_{\tau}$
and
$dim V_{\xi}=d_{\tau}\; \forall \xi$.
$U^{\l}(u)^I\,_J$ ($u\in U$) are the matrix coefficients of the operator
$U^{\l}(u)$ of the irrep $\l$
in a  (fixed)
orthonormal basis $\{{\bf v}_I\}$ of $V_{\l}$
adapted to the above decomposition. Thus
$U^{\l}(u)^{a\xi}\,_{I}=<\!U^{\l}(u){\bf v}_I,{\bf v}_{a\xi}\!>$, where
$\{{\bf v}_{a\xi}\}_{a=1,\ldots ,d_{\tau}}$ is an orthonormal basis of
$V_{\xi}$.

For scalar fields on a SS it is known
(e.g. ref.
\cite{HELGA3}) that each  $\l\in \hat{U}(\tau)$ (where $\tau$ is the trivial
representation of $K$) contains $\tau$ exactly once, so that
 $\xi_{\l}=1 \;\forall \l$.
For arbitrary fields on
arbitrary symmetric spaces $\xi_{\l}\geq 1$ and the
sum over
$\xi$ in (\ref{UNO}) is generally nontrivial.~\footnote{It is known
\cite{GODEMENT,KNAPP}
that $\xi_{\l}$ is bounded by $d_{\tau}$, giving an estimate
independent of
$\l$. There is a formula of Kostant for the multiplicity of $\tau$ in
$\l|_K$,
which, however, is too complicated for actual computation.}
In the case of
$U=Spin(N+1),SU(N+1)$
and $K=Spin(N),S(U(1)\times U(N))$, respectively,
it can be proved (see e.g.
\cite{TOM}) that
any $\tau\in\hat{K}$ appears in any $\l\in \hat{U}$ at most once.

Let $dx$ be the invariant measure on $U/K$ (induced by the Riemannian
structure) normalized by $\int_{U/K}dx=1$. Then
the ``Fourier coefficients" $\psi^I_{\l \xi}$ in (\ref{UNO})
can be determined by the formula
\beq
\label{IF45}
\psi^I_{\l \xi}={d_{\l}\over d_{\tau}}\int_{U/K}\sum_{a=1}^{d_{\tau}}
\overline{U^{\l}(\s^{-1}(x))^{a\xi}\,_I } \,\psi^a(x) dx,
\eeq
where we have used  the Schur relations for the compact group
$U$. The matrix coefficients in eq. (\ref{UNO}) satisfy the
orthogonality condition on $U/K$:
\beq
\label{ORTHGH}
\int_{U/K}\sum_{a}\overline{U^{\l}(\s^{-1}(x))^{a\xi}\,_I }\,U^{{\l}'}
(\s^{-1}(x)) ^{a{\xi}'}\,_{I'}\,dx={d_{\tau}\over d_{\l}}\,
\d_{\l{\l}'}\,\d_{II'}\,\d_{\xi{\xi}'}.
\eeq

It is well known that the Levi-Civita connection of any $U$-invariant metric
on the symmetric space $U/K$ coincides
 with the canonical connection of the second kind \cite{NOM}.
It follows from this that the
matrix-valued functions on $U/K$ given by $x\rightarrow
U^{\l}(\s^{-1}(x))^{a\xi}\,_I$ are eigenfunctions of the vector-bundle
Laplacian $L_{U/K}$ with eigenvalues~\footnote{The Laplacian $L_{U/K}$
is the $U$-invariant
differential operator on $L^2(U/K,E^{\tau})$ induced by the element
$\sum_iY_i^2$ of the universal enveloping algebra of ${\cal U}$, where
$\{Y_i\}$ is an orthonormal basis of ${\cal P}$. It is
defined by  $L_{U/K}\bfpsi = \sum_i
\nabla_{Y_i^{\ast}}\nabla_{Y_i^{\ast}}\bfpsi$ (cf. (\ref{CD34})).
Eq. (\ref{EIGENVA})
follows from the relation $L_{U/K}=\O_U-\O_K$, where $\O_U=\sum_iT_i^2$ and
$\O_K=\sum_i X_i^2$ are the Casimir elements of $U$ and $K$ ($\{T_i\}$ and
$\{X_i\}$ are orthonormal bases of ${\cal U}$ and ${\cal K}$,
respectively).}
\beq
\label{EIGENVA}
-\o_{\l}=C_2(\tau)-C_2(\l),
\eeq
where
$C_2(\m)$ denotes the (second order) Casimir number of an irrep $\m$. The
degeneracy
of $-\o_{\l}$ is $d_{\l}\,\xi_{\l}$. [We are not considering here
the additional degeneracy
coming from
the fact that different $\l$'s in $\hat{U}(\tau)$
can have the same $C_2(\l)$. If the
sum over
$\l$ in eq. (\ref{DUE}) below is meant to be a sum over {\em
different}
eigenvalues, then the true degeneracy of the eigenvalue $-\o$ is $\sum_{\l\in
A_{\o}}d_{\l}\xi_{\l}$, where $A_{\o}$ is defined as the
set of all
$\l$'s in $\hat{U}(\tau)$ with Casimir value
$C_2(\l)=\o+C_2(\tau)$.]

Consider now the heat kernel $K(x,y,t)$
of the vector-bundle Laplacian acting on
$L^2(U/K,E^{\tau})$. For $t\in {\bf R}^+$ and $x,y\in U/K$,
$K(x,y,t)$ is  an element of $Hom(E_{y},E_{x})\simeq
E_x\otimes E_{y}^{\ast}$, i.e.
a linear map from the fiber at $y$ to the fiber at
$x$.  [We recall that if $V,W$
are vector spaces we have
the isomorphism $Hom(V,W)=W\otimes V^{\ast}$, where $V^{\ast}$ is the dual
of $V$, see \cite{WARNER}
vol.I p.126.] Let $p_1$ and $p_2$ be the projections from $U/K\times U/K$
onto the first and second factor $U/K$ respectively. For a vector bundle
$E$ on $U/K$ let $p_1^{\ast}E$ and  $p_2^{\ast}E$ denote the pull-back
of $E$ relative to $p_1$ and $p_2$ respectively (cf. \cite{WALLACH} p.4).
Then
the heat kernel $K(\cdot,\cdot,t)$ is
a section of the bundle $p_1^{\ast}E^{\tau}\otimes p_2^{\ast}E^{\tau\ast}$
over
$U/K\times U/K$, where $E^{\tau\ast}$ is the
 dual bundle to $E^{\tau}$.  The heat kernel
satisfies the differential equation ($L=L_{U/K}$)
\beq
\left(-\frac{\pa}{\pa t}+L_x\right)K(x,y,t)=0,
\eeq
with the initial condition
\beq
\lim_{t\rightarrow 0^+}\int_{U/K}K(x,y,t){\bfpsi}(y)dy={\bfpsi}(x),
\eeq
for any continuous section ${\bfpsi}\in \G(E^{\tau})$.
We shall denote by $K(x,y,t)^a\,_b$ ($a,b=1,\ldots, d_{\tau}$)
the matrix representing the
endomorphism $K(x,y,t)\in Hom(E_{y},E_{x})$ in the bases
$\{{\bft}_a(y)\}$ and $\{{\bft}_a(x)\}$ of $E_{y}$ and $E_{x}$
respectively, for $x,y\in U/K\setminus {\cal C}_0$.
{}From eqs. (\ref{UNO}), (\ref{IF45}), (\ref{ORTHGH}) and (\ref{EIGENVA}) we
obtain the following
eigenfunction expansion of $K(x,y,t)^a\,_b$:
\begin{eqnarray}
K(x,y,t)^{a}\,_{b}={1\over
d_{\tau}}\sum_{\l,\xi,I}d_{\l}\,U^{\l}(\s^{-1}(x))^{a\xi}\,_I\,
\overline{U^{\l}
(\s^{-1}(y))^{b\xi}\,_I}\, e^{-t\o_{\l}}\nonumber \\
\label{ORABASTA}
={1\over  d_{\tau}}\!\sum_{\l\in\hat{U}(\tau)}\!\sum_{\xi}
d_{\l}\,U^{\l}(\s^{-1}(x)\s(y))^{a\xi}\,_{b\xi}\, e^{-t\o_{\l}},\;\;\;\;\;
a,b=1,\ldots, d_{\tau}.
\end{eqnarray}

The harmonic expansions  (\ref{UNO}) and (\ref{ORABASTA}) can also be
written in index-free  notation as follows.
For each $\l\in \hat{U}(\tau)$
let $P_{\tau}$ be the projector of $V_{\l}$ onto
 $H_{\tau}\equiv \oplus_{\xi}V_{\xi}$,
the subspace of vectors of $V_{\l}$ which transform under
$K$ according to $\tau$. $H_{\tau}$ will be identified with
$V_{\tau}\otimes {\bf C}^{\xi_{\l}}$, where $\xi_{\l}$ is the multiplicity
of $\tau$ in $\l |_K$. Define the (operator-valued)
$\tau$-spherical functions $u\rightarrow \Phi^{\l}_{\tau}(u)$
on $U$ by
\beq
\label{TTHH3}
\Phi^{\l}_{\tau}(u)=P_{\tau}U^{\l}(u)P_{\tau}.
\eeq
Each $\Phi^{\l}_{\tau}(u)$ ($u\in U$) is
regarded as a linear operator on $H_{\tau}$.
Let $\{{\bf v}_a\}$ and $\{{\bf e}_{\xi}\}$ be (fixed) orthonormal bases of
$V_{\tau}$ and ${\bf C}^{\xi_{\l}}$, respectively. Then
$\{{\bf v}_{a\xi}\}\equiv
\{{\bf v}_a\otimes {\bf e}_{\xi}\}$ is an orthonormal basis of $H_{\tau}$,
and  the matrix-valued function on $U$ given by
\beq
\label{SILVIO1}
u\rightarrow \Phi^{\l}_{\tau}(u)^{a\xi}\,_{b\xi'}\equiv
U^{\l}(u)^{a\xi}\,_{b\xi'},\;\;\;\; \l\in \hat{U}(\tau),
\;\;\;\xi,\xi'=1,\ldots
,\xi_{\l},\;\;\;a,b=1,\ldots ,d_{\tau},
\eeq
(in the notations of eq. (\ref{UNO})) is just the matrix
representing $\Phi^{\l}_{\tau}(u)$ in the basis $\{{\bf v}_{a\xi}\}$.

Let $\varphi^{\l}_{\tau}(u)$ denote the partial
trace of $\Phi^{\l}_{\tau}(u)$ with respect to ${\bf C}^{\xi_{\l}}$,
i.e. in matrix notation
\beq
\label{TRACE}
\varphi^{\l}_{\tau}(u)
^a\,_b=\sum_{\xi=1}^{\xi_{\l}}
\Phi^{\l}_{\tau}(u)^{a\xi}\,_{b\xi}.
\eeq
The operator $\varphi^{\l}_{\tau}(u)$ is regarded as
an element of $Hom(V_{\tau},V_{\tau})$ and can be expressed as
\beq
\varphi^{\l}_{\tau}(u)=d_{\tau}\int_K \phi^{\l}_{\tau}
(uk^{-1})\tau(k)dk,
\eeq
where $\phi^{\l}_{\tau}(u)=Tr\Phi^{\l}_{\tau}(u)$ is known as spherical
trace function of type $\tau$.

For $x,y\in U/K$ write $x=ux_0$, $y=vx_0$ $(u,v\in U)$, and define
$\varphi_{\l}(x,y)\in Hom(E_y,E_x)$ by
\beq
\label{SILVIO4}
\varphi_{\l}(x,y)=u\,\varphi^{\l}_{\tau}(u^{-1}v)\,
v^{-1}.
\eeq
Here $v^{-1}\in Hom(E_y,E_{x_0})$ and $u\in Hom(E_{x_0},E_x)$ are the
linear maps guaranteed by the definition of homogeneous vector bundle.
It is easy to see that $\varphi_{\l}(x,y)$ is well defined, i.e. that the
right hand side of (\ref{SILVIO4}) is invariant under $u\rightarrow uk$ and
$v\rightarrow vk'$ ($k,k'\in K$).

Eqs. (\ref{UNO}) and (\ref{ORABASTA}) can then be
rewritten in operator notation as
\beq
\label{SILVIO2}
\bfpsi(x)=\frac{1}{d_{\tau}}\sum_{\l\in\hat{U}(\tau)}d_{\l}\int_{U/K}
\varphi_{\l}(x,y)\bfpsi(y)dy,
\eeq
\beq
\label{SILVIO3}
K(x,y,t)=\frac{1}{d_{\tau}}\sum_{\l\in\hat{U}(\tau)}d_{\l}
\varphi_{\l}(x,y)e^{-t\o_{\l}}.
\eeq

Eq. (\ref{SILVIO3})  shows clearly that the heat kernel
$K(x,y,t)$
is well defined for all $x$ and $y$ in $U/K$. Moreover we immediately find
for the adjoints $\varphi_{\l}(x,y)^{\dagger}=\varphi_{\l}(y,x)$, and
similarly $K(x,y,t)^{\dagger}=K(y,x,t)$.
By choosing $u=\s(x)$ and $v=\s(y)$ in (\ref{SILVIO4}) ($x,y\in
U/K\setminus {\cal C}_0$)
we can express
$\varphi_{\l}(x,y)$ in geometric form as
\beq
\label{SILVIO5}
\varphi_{\l}(x,y)=U(x,x_0)\,
\varphi^{\l}_{\tau}(\s^{-1}(x)\s(y))\,U(x_0,y),
\eeq
where $U(x,x_0)$ (resp.  $U(x_0,y)$)
is the (vector-bundle) parallel transport operator from
$x_0$ to $x$ (resp. from $y$ to $x_0$)
along the shortest geodesic between them.

In particular for $x=y$ we have $\varphi_{\l}(x,x)={\bf 1}$ (the identity
operator on $E_x$), and
\beq
\label{DUE}
K(x,x,t)={\bf 1}\frac{1}{d_{\tau}}\sum_{\l} d_{\l}\,\xi_{\l}
\,e^{-t\o_{\l}}.
\eeq
Taking the trace of this equation and
integrating over the manifold gives the
{\em
partition function}.

Suppose now that we fix one of the two points, say $x$, at the origin
$x_0$ of
$U/K$. Then
\beq
\label{EEHK9}
K(x,t)^a\,_b\equiv K(x_0,x,t)^a\,_b=\frac{1}{d_{\tau}}
\sum_{\l,\xi}d_{\l}\,
U^{\l}(\s(x))^{a\xi}\,_{b\xi}\, e^{-t\o_{\l}},
\eeq
or, in operator notation, (cf. (\ref{SILVIO3})-(\ref{SILVIO5}))
\beq
\label{EEHK91}
K(x,t)\equiv K(x_0,x,t)=\frac{1}{d_{\tau}}\sum_{\l}d_{\l}\,
\varphi^{\l}_{\tau}(\s(x))
U(x_0,x)e^{-t\o_{\l}}.
\eeq

It is easy to find the transformation property of $K(x,t)^a\,_b$ under the
action of $K$. If $x=Exp X$ ($X\in {\cal P}_0$), it is easy to see
that $kx=Exp[Ad(k)X]$ ($k\in K$).
 Then
\beq
\s(kx)=exp[Ad(k)X]=k\,exp(X)k^{-1},\;\;\;\;\;\;\;k\in K.
\eeq
{}From this and (\ref{EEHK9}) we find
\beq
\label{IPHK}
K(kx,t)^a\,_b=\sum_{c,d}\tau(k)^a\,_c\,K(x,t)^c\,_d\,\tau(k^{-1})^d\,_b.
\eeq
The trace
\beq
Tr\left[K(x_0,x,t)U(x,x_0)\right]=
\sum_aK(x,t)^a\,_a = \frac{1}{d_{\tau}}\sum
_{\l}d_{\l}\phi^{\l}_{\tau}(\s(x)) e^{-t\o_{\l}}
\eeq
is clearly invariant under the action of
$K$ (it is a {\em zonal function}), thus it can only depend on the
coordinates along a submanifold orthogonal to the orbits of $K$. For a
symmetric space $U/K$ these submanifolds are {\em maximal tori}.
Let ${\cal A}$ be a (fixed) maximal abelian subspace of ${\cal P}$. Then
the set
$A=exp{\cal A}$ is a torus in $U$, and $Exp{\cal A}=Ax_0$ is a maximal torus
in $U/K$, i.e. a maximal, totally geodesic, flat submanifold passing
through the origin.
The dimension $l$ of $A$ is called the rank of $U/K$.
The maximal tori are all conjugate under the action of $K$ and intersect
the orbits of $K$ orthogonally.
In the so-called ``polar coordinate" decomposition, every $x$ in $U/K$
 can be written (nonuniquely) as $x=kax_0$, with $k\in K$
and $a\in A$.
More precisely, let
$M$ be the centralizer of $A$ in $K$, i.e. the set of
those elements $m$ of $K$ such that $ma=am$ for each $a\in A$.
This is a closed subgroup of $K$, and the coset space $K/M$ is
diffeomorphic to the orbits of $K$ in $U/K$. Clearly $kmax_0=kax_0$, and
we can define a map $\Psi:K/M\times A\rightarrow U/K$, by $\Psi(kM,a)=kax_0$.
The properties of this map are discussed e.g. in \cite{HELGA3} p.188. To
illustrate the main ideas, let us
suppose that $U$ is simply connected and $K$ is
connected, so that $U/K$ is simply connected. Then there is an open set
${\cal Q}_0$ in ${\cal A}$, whose closure $\overline{{\cal Q}}_0$ contains
the origin, such that the mapping
\beq
\label{PDSS}
\Psi:(kM,h)\rightarrow k(exp\,h)x_0
\eeq
maps
$K/M\times \overline{{\cal Q}}_0$ onto
$U/K$ and  gives a
bijection of $K/M\times {\cal Q}_0$ onto a certain subset $(U/K)_r$ of
$U/K$ whose complement has zero measure
(see \cite{HELGA3} Th.5.11 p.190). \footnote{
The set ${\cal Q}_0$ is any component of ${\cal A}_r$ whose closure
contains the origin, where ${\cal A}_r={\cal A}\setminus D(U,K)$ and
$D(U,K)$ is the {\em diagram} of the pair $(U,K)$, i.e. the set of all
$H\in {\cal A}$ such that $\a(H)\in \pi {\bf Z}$ for some restricted root
$\a$. $D(U,K)$ is the union of finitely many families of equispaced
hyperplanes.
The set $S_{U/K}=\Psi(K/M\times D(U,K))$
is called the {\em singular
set} in $U/K$. The {\em regular set} $(U/K)_r$ is the complement
$U/K\setminus S_{U/K}$. See \cite{HELGA0} p.271.}

Let $du$, $dk$, $dm$, and $dk_M$
be the invariant measures on $U$, $K$, $M$,
  and $K/M$  normalized by
\beq
\int_Udu=\int_Kdk=\int_Mdm=\int_{K/M}dk_M=1.
\eeq
Let $dh$ be the Euclidean measure on ${\cal A}$ induced by the
scalar product in
${\cal P}$.
Then we have the integral formula
\beq
\label{IFPD1}
\int_{U/K}f(x)dx=c\int_{K/M}\int_{{\cal Q}_0}f(k(exp\,h)x_0)J^2(h)dhdk_M,
\eeq
where the function $J^2(h)$ ($h\in {\cal Q}_0$) is
given in terms of the positive
restricted roots $\a$ of the symmetric space (with multiplicity $m_{\a}$) by
\beq
J^2(h)=\prod_{\a>0}(\sin(\a(h))^{m_{\a}},
\eeq
and the constant $c$ is determined by
\beq
\label{NC45}
c^{-1}=\int_{{\cal Q}_0}J^2(h)dh.
\eeq

We also have a polar decomposition
for $U$, namely every $u\in U$ can be written (nonuniquely) as
$k_1ak_2$, with $k_1,k_2\in K$ and $a\in A$.
Using (\ref{IFPD1}) and  the general coset space formulas ($du_K\equiv dx$)
\begin{eqnarray}
\int_Uf(u)du=\int_{U/K}\left(\int_Kf(uk)dk\right)du_K,\;\;\;\;\;\;f\in C(U),\\
\label{IFCS}
\int_Kf(k)dk=\int_{K/M}\left(\int_Mf(km)dm\right)dk_M,\;\;\;\;\;\;f\in C(K),
\end{eqnarray}
it is easy to derive the following
 integral formula related to the polar decomposition of $U$:
\beq
\label{IFPD2}
\int_Uf(u)du=c\int_{K\times {\cal Q}_0\times K}f(k_1(exp\,h)k_2)
dk_2 J^2(h)dhdk_1,
\eeq
where $c^{-1}$ is given by (\ref{NC45}) (see \cite{HELGA0} Prop, 1.19 p.385).
In particular for $f$ right-$K$-invariant (\ref{IFPD2}) reduces
to (\ref{IFPD1}) (upon using (\ref{IFCS}) for the integral in $dk_1$).

In view of the polar decomposition of $U/K$
and remembering (\ref{IPHK}), it is clear that the heat kernel
(\ref{EEHK9}) is completely determined from its restriction to
a (fixed) maximal torus $Ax_0$.

Particularly important are then the $\tau$-spherical functions
(\ref{TTHH3}) and their representative matrices (\ref{SILVIO1}).
These matrices satisfy
\beq
\Phi^{\l}_{\tau}(k_1uk_2)^{a\xi}\,_{b\xi'}=\sum_{c,d}\tau(k_1)^a\,_c\,
\Phi^{\l}_{\tau}(u)^{c\xi}\,_{d\xi'}\,\tau(k_2)^d\,_b,\;\;\;\;\;
u\in U,\;\;k_1,k_2\in K,
\eeq
and are thus determined by their restriction to $A$.

A general procedure for calculating the $\tau$-spherical functions
is the following.
Consider the Casimir operator of $U$, $\O=\sum_iT_iT_i$,
where $\{T_i\}$ is a given
orthonormal basis of ${\cal U}$. $\O$ can be regarded
 as a second-order differential
operator on $U$.
The {\em generalized radial part} $\O_r$ of $\O$,
acting on the
restrictions $\Phi^{\l}_{\tau}|_A$ is defined by
\beq
(\O \Phi^{\l}_{\tau})|_A=\O_r(\Phi^{\l}_{\tau}|_A).
\eeq
It can be shown that $\O_r$
is a well defined differential operator on $A$, whose coefficients are
operator-valued functions on $A$ acting on $\Phi^{\l}_{\tau}(a)$ both from
the left and from the right.
The explicit form of $\O_r$ acting on arbitrary $\tau$-spherical functions
was first
calculated by Harish-Chandra. It is given e.g. in ref.
\cite{WARNER} vol. II p. 277,
where one needs to analytically continue
from the noncompact symmetric space $G/K$ to the compact dual space $U/K$.

Now the Casimir operator acts as scalar multiplication on the
operator-valued functions $u\rightarrow U^{\l}(u)$, $\l\in\hat{U}$.
Therefore the functions  $\Phi^{\l}_{\tau}(a)$ are
eigenfunctions
of $\O_r$ and satisfy a certain
 differential equation on $A$. If $dimA=1$ this
differential equation
reduces to an ordinary differential equation with operator coefficients,
which,
in
principle, can  be solved explicitly  in terms of hypergeometric functions.

We shall now apply these considerations to Dirac spinors on $S^N$.

\newpage

\subsection{Spinors on $S^N$}

Let $U=Spin(N+1)$, $K=Spin(N)$, $U/K=S^N$.

\

Case 1. $N$ even.
We apply the results obtained above
to the fundamental spinor representations $\tau_+$ and
$\tau_-$ of $Spin(N)$ (cf. (\ref{FWSR1})). By using the well-known
branching rule for $Spin(N+1)\supset Spin(N)$ (see e.g. \cite{BARUT}) we
find
\beq
\label{IRREPS1}
\hat{U}(\tau_+)=\hat{U}(\tau_-)=\{\l_n\equiv (n+\frac{1}{2},\frac{1}{2},\ldots
,\frac{1}{2}), \;\;\;n=0,1,\ldots \}.
\eeq
The irreps $\l_n$ are called the {\em spinor representations} of
$Spin(N+1)$. As mentioned above, the multiplicity of $\tau_{\pm}$ in
$\l_n|_K$ is 1 for all $n$.

A Dirac spinor $\bfpsi$ is a section of the (reducible) vector bundle
$E^{\tau}=E^{\tau_+}\oplus E^{\tau_-}$ defined by $\tau=\tau_+\oplus
\tau_-$, i.e. $\bfpsi\in \G(E^{\tau})$.
We have the following relation between the iterated Dirac operator
and the spinor Laplacian on a manifold $M$:
\beq
\label{PORC11}
\slash{\!
\nabla}^2=\sum_{a=1}^N\nabla^{a}\nabla_{a}-R/4,
\eeq
where $R$ is curvature scalar of $M$. For $M=S^N$ we have $R=N(N-1)$. Thus
the eigenvalues $\l_{n,N}^2$ of $-\slash{\!
\nabla}^2$ on $S^N$ are related to the eigenvalues $\o_n$ of
$-\sum_a\nabla^a\nabla_a$ by
\beq
\label{RAL1}
\l_{n,N}^2=\o_n+N(N-1)/4.
\eeq
Now from eq. (\ref{EIGENVA})  we have for $E^{\tau_+}$
\beq
\label{RAL2}
\o_n = \o_{\l_n}=C_2(\l_n)-C_2(\tau_+).
\eeq
For $E^{\tau_-}$ we get the same eigenvalues since
$C_2(\tau_+)=C_2(\tau_-)$ (see below).

The heat kernel of $\slash{\!
\nabla}^2$ (with one point at the origin) is the direct sum
$K=K^+\oplus K^-$, where  the heat kernels $K^{\pm}$ for $E^{\tau_{\pm}}$
are given, in operator notation, by (cf. (\ref{EEHK91}))
\begin{eqnarray}
\label{TACCO1}
K^+(x,t)\equiv K^+(x_0,x,t)
={1\over \O_Nd_{\tau_+}}\sum_{n=0}^{\infty}d_{\l_n}\Phi^n_+(\s(x))
U(x_0,x)e^{-t\l_{n,N}^2} ,\\
\label{TACCO2}
K^-(x,t)\equiv K^-(x_0,x,t)
={1\over \O_Nd_{\tau_-}}\sum_{n=0}^{\infty}d_{\l_n}\Phi^n_-(\s(x))
U(x_0,x)e^{-t\l_{n,N}^2}.
\end{eqnarray}
The extra factor $1/\O_N$ occurs here because we have chosen the normalization
in which the total volume of $S^N$ is $\O_N$ (given by (\ref{OMEGAN})) in order
to compare our results with those of section 4.
The $\Phi_{+}^n$ ($\Phi^n_-$) are the $\tau_+$ ($\tau_-$)-spherical functions,
i.e. the linear
operators in $V_{\tau_+}$ ($V_{\tau_-}$) defined by
\beq
\label{OFFA2}
\Phi_+^n(u)=P_{\tau_+}U^{\l_n}(u)P_{\tau_+},\;\;\;\;\;\;\;u\in U,
\eeq
and a similar relation for $\Phi_-^n$, where $P_{\tau_{\pm}}$ are the
projectors of $V_{\l_n}$ onto the subspaces where $\l_n|_K$ is equivalent
to $\tau_+$ and $\tau_-$, respectively.
For $x,y$ nonconjugate points on $S^N$
$U(x,y)$  is the (vector-bundle)
parallel transport operator from the fibre $E_{y}$ to the fibre
$E_{x}$ along the shortest geodesic between $y$ and $x$.
Therefore for each $x\in S^N\setminus \{south \;pole\}$, the
$\Phi^{n}_{\pm}(\s(x))U(x_0,x)$ in
(\ref{TACCO1})-(\ref{TACCO2}) are linear maps from
$E_x$ to the fibre $E_{x_0}\simeq V_{\tau_{\pm}}$ at the origin (the
north pole).
This
way the  heat kernel $K(x,t)\equiv K(x_0,x,t)$
has the correct structure
(recall that $K(x,x',t)\in Hom(E_{x'},E_x)$).
It is easy to see that the functions $x\rightarrow
\Phi^{n}_{\pm}(\s(x))U(x_0,x)$ have a
well defined limit as $x$ approaches the south pole, so that the heat
kernel is regular everywhere (cf. (\ref{SILVIO5}), see also later).

We now need to compute $C_2(\l_n)$, $C_2(\tau_{\pm})$,
$d_{\l_n}$, $d_{\tau_{\pm}}$ and $\Phi_{\pm}^n$.
The Casimir number of an irrep with highest
weight $\l$ is given by Freudenthal's formula:
\beq
\label{CASI1}
C_2(\l)=(\l+\r)^2-\r^2=\l\cdot (\l+2\r),
\eeq
where $\r$ is half the sum of the positive roots of the group. $\r$ is also
given by
\beq
\r=w_1+w_2+\ldots +w_r,
\eeq
where $\{w_j\}_{j=1,\ldots ,r}$ are the fundamental weights ($r$ is the
rank of the group). The fundamental weights of $U=Spin(N+1)$ ($N$ even) are
given by (4) p.224 of \cite{BARUT}. Thus the element $\r$ for $U$ is
\beq
\r=(\frac{N-1}{2},\frac{N-3}{2}, \ldots , \frac{3}{2},\frac{1}{2}).
\eeq
Using this and (\ref{IRREPS1}) in (\ref{CASI1}) we find
\beq
\label{CASI2}
C_2(\l_n)=(n+N/2)^2-N(N-1)/8.
\eeq

The calculation of $C_2(\tau_{\pm})$ is similar. The fundamental weights of
$K=Spin(N)$ ($N$ even) are given by (5) p.224 of \cite{BARUT}. Thus the
$\r$ element for $K$ is
\beq
\r=(\frac{N}{2}\!-\!1,\frac{N}{2}\!-\!2,\ldots ,1,0).
\eeq
Using this and (\ref{FWSR1}) in (\ref{CASI1}) gives
\beq
\label{CASI3}
C_2(\tau_+)=C_2(\tau_-)=N(N-1)/8.
\eeq
Using (\ref{CASI2}) and (\ref{CASI3}) in (\ref{RAL2}) and (\ref{RAL1}) we find
\begin{eqnarray}
\o_n&=&(n+N/2)^2-N(N-1)/4,\\
\label{RAL3}
\l_{n,N}^2&=&(n+N/2)^2,
\end{eqnarray}
in complete agreement with (\ref{EIEI}).

Concerning the dimensions, these are given by the Weyl formula
\beq
d_{\l}=\prod_{\a>0}{\a\cdot (\l+\r)\over \a\cdot \r},
\eeq
where the product is over the positive roots $\a$ of the group.
For $U=Spin(N+1)$
($N$ even) the
positive roots are given by
\beq
\{{\bf e}_i\}_{i=1,\ldots ,N/2},\;\;\;\;
\{{\bf e}_i+{\bf e}_j,\;\;{\bf e}_i-{\bf e}_j\}
_{1\leq i<j\leq N/2},
\eeq
where $\{{\bf e}_i\}$ is the standard orthonormal basis of ${\bf R}^{N/2}$.
A simple calculation gives the dimensions of the spinor
representations of $Spin(N+1)$ as
\beq
\label{DIMO1}
d_{\l_n}={2^{N/2}(N+n-1)!\over n!(N-1)!},
\eeq
in agreement with (\ref{DIME1}). In a similar way we calculate
\beq
d_{\tau_+}=d_{\tau_-}=2^{N/2-1},
\eeq
as claimed in section 2. Finally let us determine $\Phi_{+}^n$.

As mentioned before, since
\beq
\label{ORCU}
\Phi^n_+(k_1uk_2)=\tau_+(k_1)\Phi^n_+(u)\tau_+(k_2),\;\;\;\;\;\;
u\in U,\;\;k_1,k_2\in K,
\eeq
it is enough to calculate the restrictions
$\Phi^n_+(a)$, $a\in A$. In our case $A$ is one-dimensional (thus $A\simeq
S^1$) and it is well known that $M$, the centralizer of $A$ in $K$, may be
identified with $Spin(N-1)$. From the relation $am=ma$ ($a\in A$, $m\in M$)
and from (\ref{ORCU}) we have
\beq
\label{CR4}
\Phi^n_+(a)\tau_+(m)=\Phi^n_+(am)=\Phi^n_+(ma)=\tau_+(m)\Phi^n_+(a),
\;\;\;\;\;m\in M,\;\;a\in A,
\eeq
i.e., the operators $\Phi^n_+(a)$, $a\in A$, commute with all the operators
of the representation $\tau_+|_M$. But from the
branching rule for $Spin(N)\supset Spin(N-1)$ we immediately find that
$\tau_+|_M = \s$, the unique fundamental spinor representation of
$Spin(N-1)$. Since $\s$ is irreducible, it follows from Schur's lemma that
$\Phi^n_+(a)$ must be proportional to the identity operator in
$V_{\tau_+}$,
\beq
\Phi^n_+(a)=f_n(a){\bf 1},\;\;\;\;\;\;\;a\in A,
\eeq
where $f_n$ is a scalar function on $A$.

Let $H$ be the element of ${\cal A}$  satisfying $\a(H)=1$, where $\a$ is
the unique positive restricted root of $U/K$.
Let us normalize the scalar product $<,>$
on ${\cal A}$ (induced from that on ${\cal P}$) so that
$<\!H,H\!>=1$.
For $a\in A$ we write $a=a_{\t}=exp(\t H)$, where $\t\in {\bf R}$.
If $d(x,x')$ denotes the  geodesic distance between
$x,x'\in S^N$, then
\beq
d(a_{\t}x_0,x_0)=d(ka_{\t}x_0,x_0)=\t,\;\;\;\;\;\;\;k\in K,
\eeq
where $x_0$ is the origin (in our case the north pole).
It is clear that
$\t$ can be taken as a coordinate on $A$ (or on $Ax_0$), and we can regard
$f_n(a_{\t})$ as a function of $\t$.  Let us denote this function by the same
symbol, i.e.
$f_n(\t)$.
We shall now prove that
\beq
\label{FINEE}
f_n(\t)=\phi_n(\t),
\eeq
where $\phi_n(\t)$ is given by (\ref{SSF1}).

In order to do this, we use the explicit formula for the generalized radial
part of the Casimir operator acting on the functions $\Phi^n_+(a)$. The
general formula in \cite{WARNER} vol. II p.277 has been specialized to the
rank-one case by Thieleker in \cite{THIELEKER} (see also \cite{WALLACH}
p.281). For $U/K=S^N$ the formula simplifies further as
there is only one
positive restricted root $\a$ (with multiplicity $m_{\a}=N\!-\!1$). Thus
let $\O_M$, $\O_K$ and $\O_U$ denote
the following
elements of the universal enveloping algebra of ${\cal U}$:
\begin{eqnarray}
\O_M&=&\sum_{i=1}^{dimM}W_i^2,\\
\O_K&=&\O_M+\sum_{i=1}^{m_{\a}}X_{\a i}^2,\\
\O_U&=&H^2+\O_K+\sum_{i=1}^{m_{\a}}Y_{\a i}^2.
\end{eqnarray}
Here $\{W_i\}$ is an orthonormal basis of ${\cal M}$ (the Lie algebra of
$M$), $\{X_{\a i}\}$, $\{Y_{\a i}\}$ are orthonormal bases of ${\cal
K}_{\a}$ and ${\cal P}_{\a}$ respectively, where
\begin{eqnarray}
{\cal K}_{\a}=\{X\in {\cal K}: (ad\,h)^2X=-\a(h)^2X,\;\;\;\forall h\in
{\cal A}\},\\
{\cal P}_{\a}=\{Y\in {\cal P}: (ad\,h)^2Y=-\a(h)^2Y,\;\;\;\forall h\in
{\cal A}\}.
\end{eqnarray}
Since ${\cal U}={\cal K}\oplus {\cal P}$, and
\beq
\label{RS1}
{\cal K}={\cal K}_{\a}\oplus {\cal M},\;\;\;\;\;
{\cal P}={\cal P}_{\a}\oplus {\cal A},
\eeq
it is clear that $\O_M$, $\O_K$ and $\O_U$ are simply
the Casimir
elements of $M$, $K$, and $U$, respectively.

If $u\rightarrow \pi(u)$ is a differentiable representation
of a compact Lie group $U$ on
a vector space $V$, we denote by $d\pi$ the corresponding
representation of the Lie algebra ${\cal U}$ on $V$. Thus
\beq
\pi(exp\,X)=exp(d\pi(X)), \;\;\;\;\;\;X\in {\cal U}.
\eeq
We use the same symbol $d\pi$ to denote the uniquely defined extension of this
representation to the universal enveloping algebra of ${\cal U}$.
In order to simplify the notation we shall however omit the symbol
$d\pi$ in the following. For instance, if $T_1,T_2$ are in the universal
enveloping algebra we shall write $T_1\pi(u)T_2$ in place of
$d\pi(T_1)\pi(u) d\pi(T_2)$. Thus in (\ref{CASIM1}) $\O_M$, $\O_K$ and
$X_{\a i}$ denote $dU^{\l}(\O_M)$, $dU^{\l}(\O_K)$ and $dU^{\l}(X_{\a i})$
respectively, whereas in (\ref{CSI1})-(\ref{CSI3}) they denote
$d\tau_+(\O_M)$, $d\tau_+(\O_K)$ and $d\tau_+(X_{\a i})$.

Now let $U=Spin(N+1)$. For $\l\in\hat{U}$ let $U^{\l}(u)$
($u\in U$) denote the operators of the irrep $\l$ on the vector space
$V_{\l}$. Consider
the operator-valued function on ${\bf R}$ defined by $\t\rightarrow
h_{\l}(\t)=U^{\l}(a_{\t})$.
Using Lemma 2 of \cite{THIELEKER} (with a proper analytic continuation to the
compact case, i.e. with $t\rightarrow i\t$)
we find that $h_{\l}(\t)$
satisfies the following differential equation
\begin{eqnarray}
\left({d^2\over d\t^2}+m_{\a}\cot\t{d\over d\t}\right)h_{\l}(\t)
+h_{\l}(\t)\O_M
\nonumber \\
-{1\over \sin^2\t}\left[(\O_M-\O_K)h_{\l}(\t)+h_{\l}
(\t)(\O_M-\O_K)\right. \nonumber \\
\left. +2\cos\t
\sum_{i=1}
^{m_{\a}}X_{\a i}h_{\l}(\t)X_{\a i}\right]\nonumber \\
\label{CASIM1}
=\O_Uh_{\l}(\t)=-C_2(\l)h_{\l}(\t).
\end{eqnarray}

Now let $\l\in\hat{U}(\tau_+)$, i.e. $\l=\l_n$, and let $\Phi^n_+(a_{\t})$
 be  the $\tau_+$-spherical
function
$P_{\tau_+}h_{\l_n}(\t)P_{\tau_+}$. By acting with $P_{\tau_+}$ both
from the left and from the right in
(\ref{CASIM1}) we obtain an equation for $\Phi^n_+(a_{\t})$.  Observe that
for $T_1$ and $T_2$ in the universal enveloping algebra of ${\cal K}$
we have (see
\cite{THIELEKER} lemma 7)
\beq
P_{\tau_+}T_1h_{\l_n}(\t)T_2P_{\tau_+}=
T_1 \Phi^n_+(a_{\t}) T_2.
\eeq
(Here $T_1,T_2$ mean $dU^{\l_n}(T_1),\; dU^{\l_n}(T_2)$ in the left-hand
side, and $d\tau_+(T_1),\;d\tau_+(T_2)$ in the right-hand side.)

It is easy to see that
\begin{eqnarray}
\label{CSI1}
\O_K \Phi^n_+(a_{\t})=\Phi^n_+(a_{\t})\O_K=-C_2(\tau_+)\Phi^n_+(a_{\t}),\\
\label{CSI2}
\O_M \Phi^n_+(a_{\t})=\Phi^n_+(a_{\t})\O_M=-C_2(\s)\Phi^n_+(a_{\t}).
\end{eqnarray}
Furthermore since $\Phi^n_+(a_{\t})$ is just a scalar operator, we have
\begin{eqnarray}
\sum_{i=1}
^{m_{\a}}X_{\a i}\Phi^n_+(a_{\t})X_{\a i}=\sum_{i=1}
^{m_{\a}}X_{\a i}X_{\a i} \,\Phi^n_+(a_{\t})\nonumber \\
\label{CSI3}
=(\O_K-\O_M)\Phi^n_+(a_{\t})=(C_2(\s)-C_2(\tau_+))\Phi^n_+(a_{\t}).
\end{eqnarray}

Using these relations and (\ref{CASIM1}) we find
the following differential equation for the scalar function
$f_n(\t)$:
\begin{eqnarray}
\left({d^2\over d\t^2}+(N-1)\cot\t{d\over d\t}\right)f_n(\t)
-C_2(\s)f_n(\t)
\nonumber \\
-{1\over \cos^2(\t/2)}\left(C_2(\tau_+)-C_2(\s)\right)f_n(\t)\nonumber \\
=-C_2(\l_n)f_n(\t). \label{CASIM2}
\end{eqnarray}

Now $C_2(\s)$ may be obtained from (\ref{CASI3}) by letting $N\rightarrow
N-1$, i.e.
\beq
\label{CASI4}
C_2(\s)=(N-1)(N-2)/8.
\eeq
Using  (\ref{CASI2}), (\ref{CASI3}) and (\ref{CASI4}) in (\ref{CASIM2}) we
obtain
\begin{eqnarray}
\left({d^2\over d\t^2}+(N-1)\cot\t{d\over d\t}\right)f_n(\t)
-{N-1\over 4\cos^2(\t/2)}f_n(\t)\nonumber \\
\label{DEFN}
=\left({(N-1)^2\over 4}-\left(n+{N\over 2}\right)^2
\right) f_n(\t).
\end{eqnarray}
This is precisely the differential equation satisfied by the functions
$\phi_n(\t)$ in (\ref{SSF1}). Obviously $\Phi^n_+(e)={\bf 1}$, and
$f_n$ must satisfy  $f_n(0)=1$. Thus (\ref{FINEE}) is proved and
$\Phi^n_+(a_{\t})=\phi_n(\t){\bf 1}$.
Notice that
 the open set ${\cal Q}_0$
(cf. subsection 5.1) is $(0,\pi)$ in this case.
Since $\Phi^n_+(a_{\pi})=0$, the functions $x\rightarrow
\Phi^n_{+}(\s(x))U(x_0,x)$ (cf. (\ref{TACCO1})) are well-defined also at
the south pole (where they vanish).
By proceeding in a similar way with
$\tau_-$ we find the same expression for $\Phi^n_-(a_{\t})$, namely
$\Phi^n_-(a_{\t})=\phi_n(\t){\bf 1}$. Using these formulas in
(\ref{TACCO1})-(\ref{TACCO2}), with $x=a_{\t}x_0$, $a_{\t}=exp(\t H)=\s(x)$,
we find
\beq
\label{HK27}
K(a_{\t}x_0,t)=U(x_0,a_{\t}x_0)f_N(\t,t),
\eeq
where $f_N(\t,t)$ is given by (\ref{HK4}).
For a generic point $x=ka_{\t}x_0\in S^N$
($k\in K$) we have $K(x,t)=U(x_0,x)f_N(\t,t)$.
Using (\ref{HK27}) and (\ref{IPHK}) we also
find $K(kax_0,t)^a\,_b=K(ax_0,t)^a\,
_b$ ($a\in A$), i.e. the matrix representing $K(x,t)\equiv K(x_0,x,t)$
in the parallel transported bases $\{{\bft}_a(x)\}$ and
$\{{\bft}_a(x_0)\}=\{{\bf v}_a\}$ (cf. subsection 5.1),
and not just its trace,
is a zonal function. The heat kernel $K(x,t)$
is well defined also at the ``antipodal manifold" $x=a_{\pi}x_0$ (the
south-pole), where it vanishes. Finally, the heat kernel $K(x,x',t)$ for
arbitrary points $x,x'$ is given by eq. (\ref{HK5}).

\

Case 2. $N$ odd. We
apply the results of subsection 5.1 to the
fundamental spinor representation $\tau$ of $K=Spin(N)$ (given by (\ref{HW2})).
By using the branching rule for \linebreak $Spin(N+1)\supset Spin(N)$ we find
\beq
\hat{U}(\tau)=\{\l_n^+,\;n=0,1,\ldots \}\;\bigcup \;\{\l_n^-,\;n=0,1,\ldots \},
\eeq
where $\l_n^+$ and $\l_n^-$ are the spinor representations of $Spin(N+1)$,
with highest weights
\beq
\l_n^{\pm}=(n+\frac{1}{2},\frac{1}{2},\ldots ,\frac{1}{2},\pm \frac{1}{2}).
\eeq

By calculating the Casimir numbers
and the dimensions of these irreps as before, we find that $C_2(\l^+_n)$ is
equal to
$C_2(\l^-_n)$ and is given by the right hand side of
(\ref{CASI2}), and similarly $d_{\l^+_n}=d_{\l_n^-}$ is given by the right
hand side of (\ref{DIMO1})  with $2^{N/2}\rightarrow 2^{(N-1)/2}$,
in agreement with our results in section 3. Also
$d_{\tau}=2^{(N-1)/2}$ and $C_2(\tau)=N(N-1)/8$.
The eigenvalues $\l_{n,N}^2$ of   $-\slash{\!\nabla}^2$
are given by (\ref{RAL1})-(\ref{RAL2}),
and we obtain again (\ref{RAL3}).

Thus the heat kernel of $\slash{\!\nabla}^2$ (with one point at the origin)
takes the form (cf. (\ref{EEHK91}))
\beq
\label{HKBASTA}
K(x,t)\equiv K(x_0,x,t)
={1\over \O_Nd_{\tau}}\sum_{n=0}^{\infty}d_{\l_n^+}
\left(\Phi^{n+}(\s(x))+\Phi^{n-}(\s(x))\right)U(x_0,x)e^{-t\l_{n,N}^2},
\eeq
where the $\tau$-spherical functions $\Phi^{n+}$ and $\Phi^{n-}$ are
defined by
\beq
\Phi^{n+}(u)=P_{\tau}U^{\l_n^{+}}(u)P_{\tau},\;\;\;\;\;u\in U,
\eeq
with an analogous relation for $\Phi^{n-}$.
[As before $P_{\tau}$ projects each representation space onto the subspace
of vectors which transform under $K$ according to $\tau$, and
$U(x_0,x)$ is the parallel spinor propagator from $x$ to $x_0$.]

In order to calculate $\Phi^{n\pm}$ notice that the operators
$\Phi^{n\pm}(a)$, $a\in A$, commute with all the $\tau(m)$,
$m\in M\simeq Spin(N-1)$ (the centralizer of $A$ in $K$).
The branching rule for $Spin(N)\supset Spin(N-1)$ gives now $\tau|_M=
\s_+\oplus \s_-$, where $\s_+$ and $\s_-$ are
the two fundamental spinor representations of $Spin(N-1)$
(cf. (\ref{CORPO})). Suppose we fix an orthonormal basis of $V_{\tau}$
adapted to the direct sum decomposition $V_{\tau}=V_{\s_+}\oplus V_{\s_-}$.
By applying Schur lemma we have
\begin{eqnarray}
\label{OPPO1}
\Phi^{n+}(a)=f_n(a){\bf 1}_+\oplus \tilde{f}_n(a){\bf 1}_-
,\;\;\;\;\;\;a\in A,\\
\label{OPPO2}
\Phi^{n-}(a)=g_n(a){\bf 1}_+\oplus \tilde{g}_n(a){\bf 1}_-
,\;\;\;\;\;\;a\in A,
\end{eqnarray}
where $f_n$, $\tilde{f}_n$, $g_n$ and  $\tilde{g}_n$ are scalar functions
on $A$ and ${\bf 1}_{\pm}$ denote the identity operators in $V_{\s_{\pm}}$.
We shall now determine these scalar functions by using the radial
part of the Casimir operator together with the Schur orthogonality
relations for the irreps of $U$.

As before, let $\a$ be the single positive restricted root of $S^N$, and
write $a\in A$ as $a=a_{\t}=exp(\t H)$, $\t\in {\bf R}$, where $H\in {\cal A}$
is chosen so that
$<\!H,H\!>=\a(H)=1$. We consider $f_n$ etc. as functions of $\t$ and denote
them by the same symbols, i.e. $f_n(\t)$ etc.

For $U=Spin(N+1)$ and $\l\in\hat{U}$ let as before $h_{\l}(\t)$ denote the
operator valued function $\t\rightarrow U^{\l}(a_{\t})$. Then
$h_{\l}(\t)$ satisfies again (\ref{CASIM1}) (by Lemma 2 of
\cite{THIELEKER}).

Now let $\l\in \hat{U}(\tau)$, i.e. $\l=\l_n^+$ or $\l_n^-$. By acting with
$P_{\tau}$ in (\ref{CASIM1}) both from the left and from the right we
obtain an equation for $\Phi^{n+}(a_{\t})$ and $\Phi^{n-}(a_{\t})$.

Since $C_2(\s_+)=C_2(\s_-)=(N-1)(N-2)/8$, we have
\beq
\label{STRANA0}
(\O_M-\O_K)\Phi^{n\pm}(a_{\t})=\Phi^{n\pm}(a_{\t})(\O_M-\O_K)=
(C_2(\tau)-C_2(\s_+))\Phi^{n\pm}(a_{\t}).
\eeq

However since $\Phi^{n\pm}(a_{\t})$ are not scalar operators, eq. (\ref{CSI3})
is no longer valid. For $p,q\in {\bf C}$ let
${\bf diag}(p,q)$ denote the operator $p{\bf
1}_+\oplus q{\bf 1}_-$ in $V_{\tau}$. Then
it is not difficult to show that
\begin{eqnarray}
\sum_{i=1}^{N-1}X_{\a i}\,{\bf diag}(p,q)X_{\a i}
&=&-(C_2(\tau)-C_2(\s_+))\,{\bf diag}(q,p)\nonumber \\
\label{STRANA}
&=&-\frac{N-1}{4}{\bf diag}(q,p).
\end{eqnarray}
Take for example $N=3$.
By writing
\beq
S^3\simeq SU(2)\simeq SU(2)\times SU(2)/SU(2)_{diag},
\eeq
we have $U=SU(2)\times SU(2)$ and $K=SU(2)_{diag}$, the diagonal subgroup
of $U$.
Let ${\cal S}$ denote the Lie algebra of
$SU(2)\simeq Spin(3)$. Then ${\cal U}={\cal S}\oplus {\cal S}$ and the
subspaces ${\cal K}$ and ${\cal P}$ in the reductive decomposition ${\cal
U}={\cal K}\oplus {\cal P}$ are easily identified as
\begin{eqnarray}
{\cal K}=\{(X,X),\;\;X\in {\cal S}\},\\
{\cal P}=\{(X,-X),\;\;X\in {\cal S}\}.
\end{eqnarray}
Take ${\cal A}={\bf R}(T_3,-T_3)\subset {\cal P}$,
where $\{T_1,T_2,T_3\}$ is a basis of ${\cal
S}$ satisfying the commutation relations
\beq
[T_1,T_2]=T_3, \;\;\;[T_2,T_3]=T_1,\;\;\;[T_3,T_1]=T_2.
\eeq
Then $(adT_3)^2T_i=-T_i$ ($i=1,2$) and comparing with (\ref{RS1}) we see
that the subspaces ${\cal K}_{\a}$ and ${\cal P}_{\a}$ are generated by
$\{(T_i,T_i), \;\;i=1,2\}$ and $\{(T_i,-T_i),\;\;i=1,2\}$, respectively.

Now let $\tau$ be the $j=1/2$ fundamental spinor representation of
$SU(2)_{diag}$.
Then
\beq
d\tau((T_3,T_3))=-\frac{i}{2}\left( \begin{array}{cc}1&0 \\  0&-1
\end{array}\right),
\eeq
and
\beq
d\tau(X_{\a 1})=
d\tau((T_1,T_1))=-\frac{i}{2}\left( \begin{array}{cc}0&1 \\
1&0
\end{array}\right),\;\;\;\;\;
d\tau(X_{\a 2})=d\tau((T_2,T_2)))=\frac{1}{2}\left( \begin{array}{cc}0&-1 \\
1&0
\end{array}\right),
\eeq
where $i=\sqrt{-1}$.
{}From this we find (omitting the symbol $d\tau$ for simplicity)
\beq
\sum_{i=1}^2 X_{\a i}{\bf diag}(p,q)X_{\a i}=-\frac{1}{2}{\bf diag}(q,p),
\eeq
which is eq. (\ref{STRANA}) for $N=3$.
The proof for $N$ (odd) $>3$ is similar and will be
omitted.

Let $\pa_{\t}\equiv d/d\t$.
Using (\ref{STRANA0}) and (\ref{STRANA}) in (\ref{CASIM1}) we get the
following set of coupled equations for the functions $f_n$, $\tilde{f}_n$:
\beq
\label{SIDAI}
\left\{ \begin{array}{ll}
\left[\pa_{\t}^2+(N\!-\!1)\cot\t\pa_{\t} -C_2(\s_+)\right]
f_{n}-\frac{N-1}{2\sin^2\t} f_n
+\frac{(N\!-\!1)\cos\t}{2\sin^2\t}\tilde{f}_n=-C_2(\l_n^+)f_n,\\
\left[\pa_{\t}^2+(N\!-\!1)\cot\t\pa_{\t} -C_2(\s_+)\right]\tilde{f}_{n}-
\frac{N-1}{2\sin^2\t} \tilde{f}_n
+\frac{(N\!-\!1)\cos\t}{2sin^2\t}f_n=-C_2(\l_n^+)\tilde{f}_n.
\end{array}
\right.
\eeq
The same set of equations (with $f_n\rightarrow g_n$ and
$\tilde{f}_n\rightarrow \tilde{g}_n$)
is obtained for the irreps $\l=\l_n^-$.

By taking the sum and the difference of the two equations in (\ref{SIDAI}),
it is then easy to see that the function $f_n(\t)+\tilde{f}_n(\t)$
satisfies the same differential equation (\ref{DEFN}) as $\phi_n(\t)$
(given by
(\ref{SSF1})), and the function $f_n(\t)-\tilde{f}_n(\t)$ satisfies the
same equation as the function $\psi_n(\t)$ given by
\beq
\label{SSF11}
\psi_n(\theta)={\psi_{n0}(\t)\over \phi_{n0}(0)}=
{n!\,\G(\frac{N}{2})\over
\G(n\!+\!\frac{N}{2})}\,\sin\frac{\t}{2}\,P_n^{(
\frac{N}{2},\frac{N}{2}-1)}(\cos\,\t),
\eeq
in the notations of section 3. [The equation for $\psi_n$ is the same as
(\ref{DEFN}) with the replacement $cos^2\t/2\rightarrow sin^2\t/2$ in the
third term of the left-hand side.]

Thus $f_n(\t)+\tilde{f}_n(\t)\propto \phi_n(\t)$,
and since $\phi_n(0)=f_n(0)=\tilde{f}_n(0)=1$ we have
\beq
f_n(\t)+\tilde{f}_n(\t)=2\phi_n(\t).
\eeq
Similarly, we have
\beq
f_n(\t)-\tilde{f}_n(\t)=2c\psi_n(\t),
\eeq
where $c$ is a (possibly complex) constant to be determined. Thus
\begin{eqnarray}
\label{LOCCO1}
f_n(\t)=\phi_n(\t)+c\psi_n(\t), \\
\label{LOCCO2}
\tilde{f}_n(\t)=\phi_n(\t)-c\psi_n(\t).
\end{eqnarray}
For the functions $g_n$ and $\tilde{g}_n$ we obtain similarly
\begin{eqnarray}
g_n(\t)=\phi_n(\t)+d\psi_n(\t), \\
\tilde{g}_n(\t)=\phi_n(\t)-d\psi_n(\t),
\end{eqnarray}
where $d$ is another constant to be determined. Notice that necessarily
$c\neq d$,  otherwise the functions $\Phi^{n+}(u)$ and $\Phi^{n-}(u)$
would coincide for all $u$ in $U$, violating the Schur orthogonality
relations for the matrix coefficients of $U^{\l_n^{\pm}}$.

It is easy to see
that $c$ and $d$ must be pure imaginary. Indeed the
$\tau$-spherical functions $\Phi^{\l}_{\tau}(u)=P_{\tau}U^{\l}(u)P_{\tau}$
satisfy
$\Phi^{\l}_{\tau}(u^{-1})=\Phi^{\l}_{\tau}(u)^{\dagger}$ (since the
operators $U^{\l}(u)$ are unitary).
At $u=a_{\t}=exp(\t H)$ this gives
$f_{n}(-\t)=\overline{f_n(\t)}$ and a similar relation for $\tilde{f}_n$.
Therefore if $f_n,\tilde{f}_n$ are real they must be even functions of $\t$,
if they are complex their real parts must be even and their imaginary
parts must be odd. Then using
(\ref{LOCCO1})-(\ref{LOCCO2}) and noting that
$\phi_n(-\t)=\phi_n(\t)$, $\psi_n(-\t)=-
\psi_n(\t)$,
we see that $\bar{c}=-c$. In a
similar way we find
$\bar{d}=-d$.

In order to determine $|c|$ and $|d|$ we write down the Schur orthogonality
relations for the matrix coefficients $U^{\l_n^+}(u)^{a\xi}\,_{b\xi}$ and
$U^{\l_n^-}(u)^{a\xi}\,_{b\xi}$.
Using the polar decomposition $u=k_1ak_2$  and the integral formula
(\ref{IFPD2}) we obtain the following three equations:
\begin{eqnarray}
\label{ORS1}
\int_0^{\pi}\left[f_n(\t)f_n(-\t)+\tilde{f}_n(\t)\tilde{f}_n(-\t)\right]
(\sin\t)^{N-1} d\t=\frac{2d_{\tau}}{d_{\l_n^+}}\frac{\O_N}{\O_{N-1}},\\
\label{ORS2}
\int_0^{\pi}\left[g_n(\t)g_n(-\t)+\tilde{g}_n(\t)\tilde{g}_n(-\t)\right]
(\sin\t)^{N-1} d\t=\frac{2d_{\tau}}{d_{\l_n^-}}\frac{\O_N}{\O_{N-1}},\\
\label{ORS3}
\int_0^{\pi}\left[f_n(\t)g_n(-\t)+\tilde{f}_n(\t)\tilde{g}_n(-\t)\right]
(\sin\t)^{N-1} d\t=0.
\end{eqnarray}
We have used here the fact that in the present normalization the constant
$c^{-1}$ in (\ref{NC45}) is given by
\beq
\int_{{\cal Q}_0}J^2(h)dh=\int_0^{\pi}(\sin \t)^{N-1}d\t=\O_N/\O_{N-1}.
\eeq

Now using the above equations to express $f_n$ etc. in terms of $\phi_n$
and $\psi_n$, and remembering that
$\phi_n$ is even and $\psi_n$ is odd,
 we can rewrite (\ref{ORS1})-(\ref{ORS3}) as
\begin{eqnarray}
\label{ORS4}
\int_0^{\pi}(\phi_n^2-c^2\psi_n^2)(\sin\t)^{N-1}d\t=\frac{2\O_N}{d_n\O_{N-1}},\\
\label{ORS5}
\int_0^{\pi}(\phi_n^2-d^2\psi_n^2)(\sin\t)^{N-1}d\t=\frac{2\O_N}{d_n\O_{N-1}},\\
\label{ORS6}
\int_0^{\pi}(\phi_n^2-cd\psi_n^2)(\sin\t)^{N-1}d\t=0,
\end{eqnarray}
where $d_n$ is given by (\ref{SDSNAA}).
Now from (\ref{NORMALI}) we have
\beq
\label{BOSTA2}
\int_0^{\pi}\phi_n^2(\t)(\sin\t)^{N-1}d\t=
\int_0^{\pi}\psi_n^2(\t)(\sin\t)^{N-1}d\t=\frac{\O_N}{d_n\O_{N-1}}.
\eeq
Using this in (\ref{ORS4})-(\ref{ORS6})
we see that $c^2=d^2=-1$ and $cd=1$,
whence
\beq
c=i=\sqrt{-1},\;\;\;\;\;\;d=-i.
\eeq

We have obtained
\begin{eqnarray}
\label{SPIT1}
f_n&=&\phi_n+i\psi_n=\tilde{g}_n,\\
\label{SPIT2}
\tilde{f}_n&=&\phi_n-i\psi_n=g_n.
\end{eqnarray}
Notice that $\Phi^{n\pm}(a_{\pi})=\pm i(-1)^n{\bf diag}(1,-1)$,
i.e.  the spherical functions do not vanish at $\t=\pi$ in this case (but
the heat kernel does).

Let us prove that the limit as $x$
approaches the south pole of the functions \linebreak
$\Phi^{n\pm}(\s(x))U(x_0,x)$
(cf. (\ref{HKBASTA})) exists (i.e. it does not depend on the direction).
Recall that the parallel transport operator $U(x_0,x)$ from $x=ExpX$
($X\in {\cal P}_0$) to the origin $x_0=NP$ (the north pole)
along the geodesic $exp(-tX)x$
($t\in [0,1]$) is the same as the linear map between the fibres $E_x$ and
$E_{x_0}$ corresponding to the transvection $exp(-X)=(\s(x))^{-1}$,
see subsection 5.1. Thus $\Phi^{n\pm}(\s(x))U(x_0,x)=
\Phi^{n\pm}(\s(x))(\s(x))^{-1}$. Now notice that the functions on $U$ given
by $u\rightarrow \Phi^{n\pm}(u)u^{-1}\in Hom(E_{ux_0},E_{x_0})$ are
invariant under $u\rightarrow uk$ ($k\in K$). Therefore the functions
$\varphi^{n+}$ and $\varphi^{n-}$ on $U/K$ given by~\footnote{
In the notations of subsection 5.1, $\varphi^{n\pm}(x)=
\varphi_{\l_n^{\pm}}(x_0,x)$, cf. eq. (\ref{SILVIO4}).}
\beq
x=ux_0\rightarrow \varphi^{n\pm}(x)\equiv \Phi^{n\pm}(u)u^{-1}\in
Hom(E_x,E_{x_0})
\eeq
are well defined for all $x\in U/K=S^N$, in particular for $x=a_{\pi}x_0=SP$
(the south pole). Since $\Phi^{n\pm}(\s(x))(\s(x))^{-1}=
\varphi^{n\pm}(x)$ for $x\in S^N\setminus \{ SP\}$,
we have
$$\lim_{x\rightarrow SP}\Phi^{n\pm}(\s(x))U(x_0,x)
=\varphi^{n\pm}(SP),$$
and our claim is proved.

 Now since $K$ fixes the south pole, we have a representation
$k\rightarrow \tilde{\tau}(k)$ of $K$ on the fibre $E_{SP}$ at the south pole.
If $u$ is any element of $U$ such that $ux_0=SP$, then $u^{-1}Ku=K$ and
\beq
\label{TAUTILDE}
\tilde{\tau}(k)=u\,\tau(u^{-1}ku)u^{-1},\;\;\;\;\;k\in K.
\eeq
It is easy to see that the
linear maps $\varphi^{n\pm}(SP)\in Hom(E_{SP},E_{NP})$
intertwine $\tilde{\tau}$ and $\tau$, i.e.
\beq
\varphi^{n\pm}(SP)\tilde{\tau}(k)=\tau(k)\varphi^{n\pm}(SP),
\;\;\;\;\;\forall k\in K.
\eeq
Indeed let $u$ be any element of $U$ such that $ux_0=SP$ and let $k\in K$.
Then
\begin{eqnarray}
\varphi^{n\pm}(SP)\,\tilde{\tau}(k)
=\Phi^{n\pm}(u)u^{-1}u\,\tau(u^{-1}ku)u^{-1}
\nonumber \\
=\Phi^{n\pm}(ku)u^{-1}
=\tau(k)\Phi^{n\pm}(u)u^{-1}\nonumber \\
=\tau(k)\varphi^{n\pm}(SP),
\end{eqnarray}
as claimed.

Since $\Phi^{n\pm}(a_{\pi})=
\pm i(-1)^n{\bf diag}(1,-1)$, it is clear that the $\varphi^{n\pm}(SP)$ are
nontrivial intertwining operators, so that $\tilde{\tau}$ is equivalent
to $\tau$. This is always true in the odd dimensional case, namely
one can show [using (\ref{TAUTILDE})]
that for $N$
odd, given {\em any} $\tau\in \hat{K}$,
the representation $\tilde{\tau}$ of $K$
on the fibre $E_{SP}$ of the homogeneous vector bundle $E^{\tau}$ over
$S^N$ is always equivalent to $\tau$.

For $N$ even one can show that if $\tau$ is the irreducible representation
of $K=Spin(N)$ with highest weight $(p_1,p_2,\ldots ,p_{N/2})$, then
$\tilde{\tau}=(p_1,p_2,\ldots ,p_{{N/2}-1},-p_{N/2})$. For example in our
spinor case we have $\tilde{\tau}_+=\tau_-$ and $\tilde{\tau}_-=\tau_+$
(cf. (\ref{FWSR1})). If $\Phi^n_+$ is the $\tau_+$-spherical function
(\ref{OFFA2}), the limit $\varphi^n_+(SP)=\lim_{x\rightarrow
SP}\Phi^n_+(\s(x))U(x_0,x)$ is again well defined and $\varphi^n_+(SP)$
intertwines $\tilde{\tau}_+$ with $\tau_+$ (same proof as above). However
since $\tilde{\tau}_+=\tau_-\not\sim \tau_+$, it follows from Schur lemma that
$\varphi^n_+(SP)=0$. This explains our previous result that
$\Phi^n_+(a_{\pi})=0$ for $N$ even.

Using (\ref{SPIT1})-(\ref{SPIT2}) in (\ref{OPPO1})-(\ref{OPPO2}) and then in
(\ref{HKBASTA}) we find that the heat kernel $K(a_{\t}x_0,t)$, $a_{\t}
=exp(\t H)$, is
given again by (\ref{HK27}), and $K(x,x',t)$ is given by (\ref{HK5}).
Thus we find complete agreement with our results of section 4.

\newpage
\subsection{Harmonic analysis on noncompact symmetric spaces}

Let $G$ be a noncompact semisimple Lie group with finite center, $K$ a
maximal compact subgroup, and $G/K$ the associated Riemannian symmetric
space of the noncompact type.
The results obtained in subsection 5.1 can be generalized to homogeneous
vector bundles $E^{\tau}$ over $G/K$ ($\tau\in\hat{K}$).
Harmonic analysis on $G/K$ is well
understood in the case of scalars. For vector bundles $L^2$-harmonic
analysis on $G/K$ may be reduced to $L^2$-harmonic analysis on the group $G$,
by letting $L^2(G/K,E^{\tau})$ sit in $L^2(G)$ in a natural way.

In the noncompact case  $\hat{G}$ (the set of equivalence classes of
irreducible unitary representations of $G$) is not completely known, in
general.
 Fortunately, we do not need to know all of
$\hat{G}$ in order to do harmonic analysis. Indeed a representation $\l$ in
$\hat{G}$ may contain a given $\tau\in \hat{K}$ and still not appear in the
direct integral decomposition (over $\hat{G}$) of the induced
representation $\pi_{\tau}$ of $G$ on $L^2(G/K,E^{\tau})$.
In other words, the measure $d\m(\l)$ in this decomposition (known as the
{\em Plancherel measure}) has support in a proper subset of $\hat{G}$.
The main problem
 is then the explicit determination of the Plancherel measure and of the
{\em tempered} representations (i.e. those
representations  with non-zero Plancherel measure).

For semisimple Lie groups this problem has been solved completely by
Harish-Chandra (see e.g. \cite{KNAPP}). Thus, harmonic analysis for bundles
on $G/K$ is, in principle, known. Notice that this is only true for $K$
compact, i.e. in the Riemannian case.
For (pseudo-Riemannian) symmetric spaces $G/H$, with $H$
noncompact, we may have a contribution to the analysis on $L^2(G/H)$ from
representations of $G$ with zero Plancherel measure in $\hat{G}$ (the
so-called ``complementary series").

We shall not discuss these matters further here. The important point for us
is that the heat kernel of the Laplacian acting on $L^2(G/K,E^{\tau})$
may be expanded in terms of the $\tau$-spherical functions in a way which
is similar to (\ref{EEHK91}) with $\sum_{\l\in\hat{U}(\tau)}d_{\l}
\rightarrow
\int_{\hat{G}(\tau)}d\m(\l)$, i.e.
\beq
\label{NCHK78}
K(x,t)\equiv K(x_0,x,t)={1\over d_{\tau}}\int_{\hat{G}(\tau)}
\varphi^{\l}_{\tau}(\s(x))
U(x_0,x)\,e^{-t\o_{\l}}d\m(\l).
\eeq
The notations here are as follows. The $\tau$-spherical functions are the
(operator-valued) functions $g\rightarrow \Phi^{\l}_{\tau}(g)$ on $G$
defined by
\beq
\Phi^{\l}_{\tau}(g)=P_{\tau}U^{\l}(g)P_{\tau},\;\;\;\;\;\;g\in G,
\eeq
where  $U^{\l}(g)$ denote the
operators of the representation $\l\in\hat{G}(\tau)$ (this set has the same
meaning as for the compact group $U$) in a Hilbert space $H_{\l}$, and
$P_{\tau}$ is the projector of $H_{\l}$ onto
$H_{\tau}=\oplus_{\xi=1}^{\xi_{\l}}V_{\xi}\simeq V_{\tau}\otimes {\bf
C}^{\xi_{\l}}$,
the subspace of vectors of $H_{\l}$ which
transform
under $K$ according to $\tau$. As before, $\xi_{\l}$ denotes
the multiplicity of $\tau$ in ${\l}|_K$, and
$\varphi^{\l}_{\tau}(g)$ denotes the partial
trace of $\Phi^{\l}_{\tau}(g)$ with respect to ${\bf C}^{\xi_{\l}}$ (cf.
(\ref{TRACE})).
The operator $\Phi^{\l}_{\tau}(g)$  is regarded as an element of
$Hom(H_{\tau},H_{\tau})$ and  $\varphi^{\l}_{\tau}(g)$ as an
element of  $Hom(V_{\tau},V_{\tau})$.

We write the reductive (Cartan) decomposition of the Lie algebra
${\cal G}$ of $G$ as ${\cal G}={\cal K}\oplus {\cal P}$, where ${\cal K}$ is
the Lie algebra of $K$ and ${\cal P}$ is a complementary
$Ad(K)$-invariant subspace  which we choose as ${\cal K}^{\perp}$,
the orthogonal
complement of ${\cal K}$ in ${\cal G}$  with respect to the Killing form $B$:
\beq
B(X,Y)=Tr\,ad(X)ad(Y),\;\;\;\;\;\;X,Y\in{\cal G}.
\eeq

We identify ${\cal P}$ with $T_{x_0}(G/K)$ and
for $x=ExpX\in G/K$ ($X\in {\cal P}$) we write $\s(x)=exp(X)$,
where as before $Exp$ and $exp$ denote the exponential mappings in
$G/K$ and $G$, respectively.

  As in the compact case $U(x_0,x)$
denotes the (vector-bundle) parallel transport operator from
$x$ to $x_0$ along the geodesic between them. Since there are
no points conjugate to $x_0$, $U(x_0,x)$ is well defined for any $x$ in
$G/K$. If $x=ExpX$ then $U(x_0,x)=exp(-X)=(\s(x))^{-1}$ as a
linear map from the fibre $E_x$ to the fibre $E_{x_0}\simeq V_{\tau}$.

It is well known that the Killing form
is positive definite on ${\cal P}$ and negative definite on ${\cal K}$. If
$\{T_i\}$ is a basis of ${\cal G}$, let $g_{ij}=B(T_i,T_j)$, and let
$g^{ij}$ denote the inverse of the matrix $g_{ij}$. The Casimir operator of
$G$ is the following element of the universal enveloping algebra of ${\cal G}$:
\beq
\O_G=\sum_{i,j=1}^{dimG}g^{ij}T_iT_j.
\eeq
Let $\{X_i\}$ be a basis of ${\cal K}$, orthonormal with respect to $-B$,
and let $\{Y_i\}$ be a basis of ${\cal P}$, orthonormal with respect to
$B$. Then
\beq
\O_G=\sum_{i=1}^{dim{\cal P}}Y_i^2-\sum_{i=1}^{dim{\cal K}}X_i^2.
\eeq
Viewed as a differential operator   on $\G(E^{\tau})$ $\sum Y_i^2$ induces the
vector-bundle Laplacian $L_{G/K}$,
and as before $\sum X_i^2=\O_K$, the (negative definite)
Casimir element of $K$. Thus
\beq
\label{RBLAC}
L_{G/K}=\O_G+\O_K.
\eeq
The eigenvalues $-\o_{\l}$ of $L_{G/K}$ acting on
the vector-valued functions $g\rightarrow P_{\tau}U^{\l}(g^{-1})v$
($v\in H_{\l}$, $\l\in\hat{G}(\tau)$)
 may then be related to the Casimir values of $\l$ and $\tau$.
Since
$\O_G$ is in the center of the universal enveloping algebra of ${\cal G}$, it
acts as a scalar in $H_{\l}$. Denoting this scalar by
$-C_2(\l)$, we have from  (\ref{RBLAC})
\beq
\label{PORC13}
-\o_{\l}=-C_2(\l)-C_2(\tau),
\eeq
to be compared to the compact relation (\ref{EIGENVA}). There is a
well-known duality between the noncompact and the compact symmetric spaces.
Consider the subspace ${\cal U}={\cal K}\oplus i{\cal P}$ of the
complexification ${\cal G}^{\bf C}$
of ${\cal G}$, where $i=\sqrt{-1}$. The Lie algebra
${\cal U}$ is called a {\em compact
form} of ${\cal G}$. It is a compact Lie algebra since the Killing form
$B$ is negative definite on
${\cal U}$.  Let $U$ denote a
simply connected compact Lie group with Lie algebra ${\cal U}$. Then $U/K$
is the compact symmetric space which is {\em dual} to $G/K$. It is clear
that the radial parts of the Casimir operators
on $G$ and $U$ (acting on $\tau$-spherical functions) will be related by
analytic continuation through
$h\rightarrow ih$, for $h\in {\cal A}$. This amounts
to changing trigonometric with hyperbolic functions.

\subsection{Spinors on $H^N$}

Let $G=Spin(N,1)$, $K=Spin(N)$, $G/K=H^N$.

\

Case 1. $N$ even. In this case the Plancherel measure in $\hat{G}$
has both a discrete
and a continuous part. The tempered representations are the discrete series
and the principal unitary series. Let $\tau=\tau_+\oplus \tau_-$ (same
notations as in subsection 5.2). From the branching rule for
$Spin(N,1)\supset Spin(N)$ we find that no discrete series contain
$\tau$ (see \cite{CAMPO3}). Let $U_{(i\l,\s)}$ denote
the unitary principal series representation
labelled by $\l\in {\bf R}$ and $\s\in\hat{M}$, where  $M=Spin(N-1)$ is the
centralizer of $A$ in $K$ and $G = KAN$ is an
Iwasawa decomposition of $G$ \cite{WALLACH,WARNER}. This is the representation
of $G$ induced by  the representation $D$
of a minimal parabolic subgroup $MAN$ given by
\beq
D(man) = \sigma(m) e^{i\l(\log a)}.
\eeq
By Frobenius Reciprocity $U_{(i\l,\s)}|_K$ contains $\tau$ if and only if
$\tau|_M$ contains $\s$. Thus
we find that the unitary
principal series containing both $\tau_+$ and $\tau_-$ are
the $U_{(i\l,\s)}$ with
$\s=(\frac{1}{2},\ldots ,\frac{1}{2})$, the fundamental spinor
representation of $M$.  The Plancherel measure for this principal series
is (see \cite{CAMPO3})
\beq
\label{PLAN1}
d\m(U_{(i\l,\s)})={2^{N-2}\over \pi\O_{N-1}}d_{\s}|C(\l)|^{-2}d\l,
\eeq
\beq
\label{PLAN2}
C(\l)={2^{N-2}\G(N/2)\over \sqrt{\pi}}{\G(i\l+\frac{1}{2})\over
\G(i\l+\frac{N}{2})},
\eeq
where $d_{\s}=2^{N/2-1}$ is the dimension of $\s$ and $\O_{N-1}$ is the
volume of $S^{N-1}$ (cf. (\ref{OMEGAN})).
As in the scalar case, the spinor Harish-Chandra function $C(\l)$ is
determined by the asymptotic form at infinity of the
$\tau$-spherical functions $\Phi^{\l}_+$ and  $\Phi^{\l}_-$,
where $\Phi^{\l}_{{\pm}}=\linebreak
P_{\tau_{\pm}}U_{(i\l,\s)}P_{\tau_{\pm}}$.
The radial part of the Casimir operator acting on the restrictions
$\Phi^{\l}_{{\pm}}(a_y)$,\linebreak $a_y=exp(y H)\in A$, may be obtained from
(\ref{CASIM1}) by letting $\t\rightarrow iy$ and applying $P_{\tau_+}$ (or
$P_{\tau_-}$) both from the left and from the right.
Again
$\Phi^{\l}_{+}(a_y)=\phi_{\l}(y){\bf 1}$, where $\phi_{\l}$ is a scalar
function on $A$.
The Casimir value $-C_2(\l)$ for $U_{(i\l,\s)}$ may be easily calculated
and is given by
\beq
-C_2(\l)=-\l^2-N(N-1)/8.
\eeq
It may be obtained from (\ref{CASI2}) by letting $n\rightarrow -i\l-N/2$.
Using (\ref{PORC11}), (\ref{CASI3}) and (\ref{PORC13}) we find that
the eigenvalues of ${\slash{\!
\nabla}}^2$ are given by
\beq
{\slash{\!
\nabla}}^2_{\l}=-\o_{\l}-R/4=-\l^2,\;\;\;\;\;\;\l\in {\bf R}.
\eeq

The differential equation
satisfied by the scalar function
$\phi_{\l}(y)$ is obtained from (\ref{DEFN}) by letting
$\t\rightarrow iy$ and $n\rightarrow -i\l-N/2$. This may be put in
hypergeometric form and we find
\beq
\label{UUGG}
\phi_{\l}(y)=\cosh\frac{y}{2}\,F(i\l+\frac{N}{2},-i\l+\frac{N}{2},\frac{N}{2},
-\sinh^2\frac{y}{2}).
\eeq

Similarly we find $\Phi^{\l}_{-}(a_y)=\phi_{\l}(y){\bf 1}$.
Using the above formulas in (\ref{NCHK78}) we find for the heat kernel
$K=K^+\oplus K^-$,
\beq
\label{NCHK99}
K^+(x,t)=K^-(x,t)=U(x_0,x)\,{2^{N-3}\G(N/2)\over
\pi^{N/2+1}}\int_0^{+\infty}\phi_{\l}(y) e^{-t\l^2}|C(\l)|^{-2}d\l,
\eeq
where $y=d(x_0,x)$. This agrees with (\ref{HK8})-(\ref{HK9}).

\

Case 2. $N$ odd. In this case the group $G=Spin(N,1)$ has no discrete
series and the
Plancherel measure is purely continuous.
Let $\tau$ be the irrep of $K$ given by (\ref{HW2}). By
Frobenius Reciprocity we find that the unitary principal series
representations containing
$\tau$ are $U_{(i\l,\s_+)}$ and $U_{(i\l,\s_-)}$,
where $\s_{\pm}=(\frac{1}{2},\ldots ,\frac{1}{2},\pm\frac{1}{2})$ are the two
fundamental spinor
representations of $M=Spin(N-1)$ (cf. (\ref{CORPO})).
The Plancherel measure $d\m(U_{(i\l,\s_+)})=d\m(U_{(i\l,\s_-)})$
is given again by
(\ref{PLAN1}) (with $d_{\s}\rightarrow d_{\s_+}=2^{(N-3)/2}$)
and (\ref{PLAN2}).

Since $\tau|_M=\s_+\oplus
\s_-$, the $\tau$-spherical functions
$\Phi^{\l\pm}(g)\equiv P_{\tau}U_{(i\l,\s_{\pm})}(g)
P_{\tau}$ at $g=a_y=exp(yH)\in A$ are given by (cf.
(\ref{OPPO1})-(\ref{OPPO2}))
\begin{eqnarray}
\label{APPO1}
\Phi^{\l+}(a_y)={\bf diag}(f_{\l}(y),\tilde{f}_{\l}(y)),\\
\label{APPO2}
\Phi^{\l-}(a_y)={\bf diag}
(g_{\l}(y),\tilde{g}_{\l}(y)).
\end{eqnarray}
The scalar functions
$f_{\l}$, $\tilde{f}_{\l}$, $g_{\l}$ and  $\tilde{g}_{\l}$
may be obtained respectively from $f_n$, $\tilde{f}_n$, $g_n$ and
$\tilde{g}_n$  in (\ref{SPIT1})-(\ref{SPIT2}) by letting $n\rightarrow
-i\l-N/2$ and changing the argument $\t\rightarrow iy$. Thus
\begin{eqnarray}
\label{SPUT1}
f_{\l}&=&\phi_{\l}+i\psi_{\l}=\tilde{g}_{\l},\\
\label{SPUT2}
\tilde{f}_{\l}&=&\phi_{\l}-i\psi_{\l}=g_{\l},
\end{eqnarray}
where $\phi_{\l}(y)$ is given by (\ref{UUGG}) and
\beq
\psi_{\l}(y)
=\frac{2\l}{N}\sinh\frac{y}{2}\,
F(i\l+\frac{N}{2},-i\l+\frac{N}{2},\frac{N}{2}+1,
-\sinh^2\frac{y}{2}).
\eeq

Using the above  equations in (\ref{NCHK78}) we
find that the heat kernel $K(x,t)$ is given by the right-hand side of
(\ref{NCHK99}), again in agreement with our results in section 4.

\newpage

\begin{flushleft}
{\Large {\bf Acknowledgments}}
\end{flushleft}
The authors would like to thank Prof. Andrzej Trautman
for useful discussions and for pointing out
Ref. [17]. The work of A.H. was
supported in part by the Schweizerisher Nationalfonds.

\end{document}